\documentclass[graybox]{svmult}

\usepackage{mathptmx}       %
\usepackage{helvet}         %
\usepackage{courier}        %
\usepackage{type1cm}        %
\usepackage{makeidx}         %
\usepackage{graphicx}        %
\usepackage{multicol}        %
\usepackage[bottom]{footmisc}%

\usepackage{amsmath}
\usepackage{amssymb}

\makeindex             %

\begin{document}

\title*{Analyzing and biasing simulations with PLUMED}
\author{Giovanni Bussi and Gareth A. Tribello}
\institute{Giovanni Bussi \at Scuola Internazionale Superiore di Studi Avanzati, Trieste, Italy, \email{bussi@sissa.it}
\and Gareth A. Tribello \at Atomistic Simulation Centre, School of Mathematics and Physics, Queen's University Belfast, Belfast, BT7 1NN, United Kingdom, \email{g.tribello@qub.ac.uk}}

\maketitle

\abstract{
This chapter discusses how the PLUMED plugin for molecular dynamics can be used to analyze and bias molecular dynamics trajectories.  The chapter begins by introducing the notion of a collective variable and by then explaining how the free energy can be computed as a function of one or more collective variables.  A number of practical issues mostly around periodic boundary conditions that arise when these types of calculations are performed using PLUMED are then discussed.  Later parts of the chapter discuss how PLUMED can be used to perform enhanced sampling simulations that introduce simulation biases or multiple replicas of the system and Monte Carlo exchanges between these replicas.  This section is then followed by a discussion on how free-energy surfaces and associated error bars can be extracted from such simulations by using weighted histogram and block averaging techniques.  
}

\section{Introduction}

The chapters in sections I to IV will have given you some sense of the broad range of methods and techniques that have been used to simulate biomolecular processes.  The aim of this chapter is not to introduce more techniques but rather to focus on how these techniques can be employed in practice.  We will do so by explaining how a particular piece of software, PLUMED
\cite{bonomi2009plumed,tribello2014plumed}, can be used to run and analyze many of the types of simulation that are discussed in section II.  Note \ref{note-plumed-versions} discusses the various different versions of PLUMED. The most important thing to know at the outset, however, is that PLUMED is not a molecular dynamics (MD) or Monte Carlo code.  It is instead designed to complement MD codes such as GROMACS~\cite{abraham2015gromacs},
LAMMPS~\cite{plimpton1995fast},
DL\_POLY~\cite{todorov2006dl_poly_3},
CP2K~\cite{hutter2014cp2k},
AMBER~\cite{amber},
and OpenMM~\cite{eastman2017openmm}.  PLUMED does this in two ways:

\begin{enumerate}
\item It can be used to post-process the molecular dynamics trajectories that are generated by these code.
\item It can serve as a plugin to these MD codes and thus allow the user to add the additional biasing forces that are required for the enhanced sampling methods, such as umbrella sampling or metadynamics, that are described in Section II.
\end{enumerate}

\begin{figure}
\centering
\includegraphics[width=0.9\textwidth]{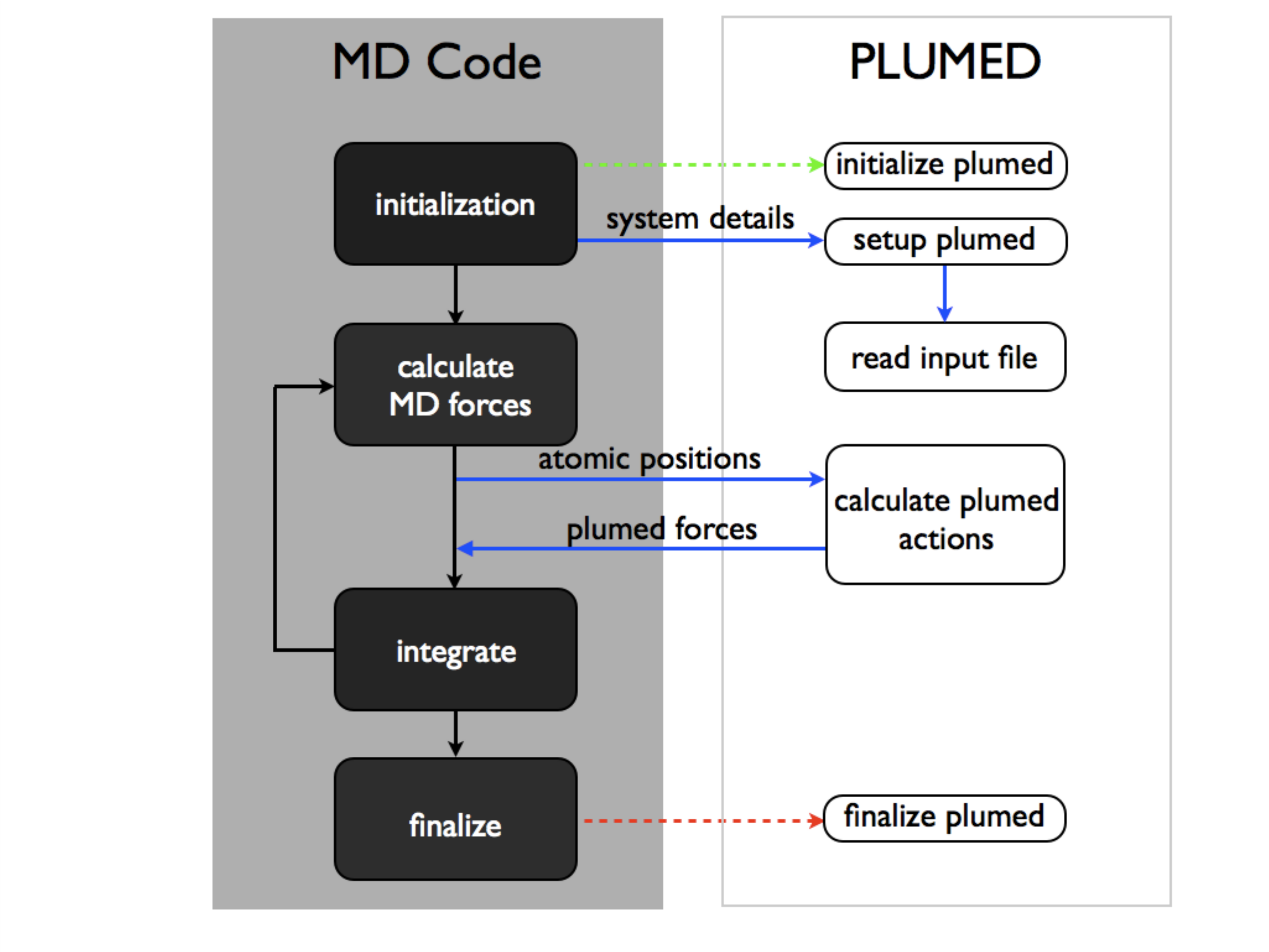}
\caption{Schematic representation of the interface between PLUMED and an MD engine.  The green arrow indicates that the function called should create the PLUMED object and the red arrow indicates that the function called should delete the PLUMED object}
\label{fig:plumed-interface}
\end{figure}

The manner in which PLUMED is plugged into an MD code is illustrated in figure~\ref{fig:plumed-interface}.  As you can see PLUMED is called during initialization and its input file is read in at that time.  It is then called during each run through of the main loop of the MD code just after the forces that describe the interactions between the atoms are calculated.
Calling PLUMED at these points allows the plugin to do any analysis that is required and also allows any forces due to bias potentials that are calculated by PLUMED to be returned from PLUMED to the MD code so that they can be incorporated when the equations of motion are integrated. PLUMED is not the only piece of software that interacts with other MD codes in this manner. Two other notable examples are the COLVARS package \cite{fiorin2013using}, which is also reviewed in this book, and the recently published program SSAGES \cite{sidky2018ssages}.

When the PLUMED package is used to post process trajectories a program that is part of PLUMED and that is called \texttt{driver} is employed. When PLUMED \texttt{driver} 
is used the trajectory is not generated by integrating the equations of motion.  The trajectory is instead read from disk so the forces calculated within PLUMED thus play no role in the systems' dynamics. PLUMED \texttt{driver} is, nevertheless, useful as the PLUMED code contains numerous analysis tools that can also be used to post process trajectories.

PLUMED is written using C++ and the object-oriented paradigm is heavily used in the design of the code.  The consequence of this is that the code contains a set of core modules that look after the communication with the MD codes and various other mission critical features.  This code is maintained by a small cadre of core developers.  Additional functionalities can be built on this core and contributed by any user in the community.  In fact, developers who wish to contribute to the project in this way often only need to contribute one single file that contains class and method definitions, the code itself and the sections of the manual that describe the new functionality as well as files for a regression test that can be used to ensure that the new functionality continues to work if changes are made elsewhere.
The fact that it is relatively easy to extend PLUMED ensures that the code, the associated website and the various user meetings provide a forum that developers can use to share new techniques and methods with the scientific community. 
In fact, in PLUMED v2.4, a modular framework was introduced that allows developers to contribute groups
of functionalities that are logically connected.  So far Omar Valsson (VES module, for variational enhanced sampling), Glen Hocky and Andrew White (EDS module, for experiment-directed simulations) and Haochuan Chen and Haohao Fu (DRR module, for extended-system adaptive biasing force),
who are all non-core developers, have used this model to contribute modules to the code base. We expect, however, that the number of contributed modules will increase in the future.

The size and scope of the PLUMED code ensures that we cannot describe everything that it can do in this single chapter.  For those who are interested we would recommend reading the original article
\cite{tribello2014plumed}, the code's online manual, the many tutorials included in the manual,  and some of the considerable number of papers that
describe simulations done or analyzed with PLUMED.
What we will do in the following is quickly summarize the theory behind some of the methods that are implemented in PLUMED.  We will also provide relevant examples of PLUMED input files that can be used for these types of calculations.
Our focus in these sections is on describing how free energies are estimated from these types of calculations, how results from multiple replicas can be combined, and how suitable error bars on these estimates are determined.

\section{Collective variables}
\label{sec:cvs}

Complex biochemical reactions or conformational changes are often interpreted in terms of free-energy surfaces that are
computed as a function of a small number of collective variables (CVs). 
These free energy surfaces, which are often referred to as potentials of mean force, provide a coarse grained representation of the energy landscape for the system, which can in turn provide useful insights into the behavior of these biochemical systems.  To see why consider the following examples:
\begin{itemize}
\item We know that nucleosides have multiple conformations and that inter-conversion between these conformers involves rotation around the glycosidic bond.  A free-energy surface for such a system as a function of the torsional angle $\chi$ around this bond is thus useful as it provides us with information on the relative free energies of the conformations and an indication of the height of the barrier for inter-conversion
(see, e.g., Ref.~\cite{gil2015enhanced}).

\item We know that the secondary and tertiary structures of many proteins determine their function and that it is important to understand the mechanism by which proteins fold.  If we run molecular dynamics simulations we can understand something about this folding process by extracting the free-energy surface along a coordinate that counts the number of native contacts that are present (see, e.g., \cite{best2013native,camilloni2014statistical}).

\item We know that cells are constantly exchanging water and ions across their membrane.  To extract information on the mechanisms for these processes we can run molecular dynamics simulations and extract a free-energy surface that describes how the free energy of the ion changes as the ion moves through an ion channel (see, e.g., \cite{zhang2011combined}).

\item We know that the behavior of biomolecules changes when their protonation state changes.  To extract information on the likely protonation state of a biomolecule we might, therefore, calculate how the free energy of the molecule changes as the distance between the hydrogen atoms and the various protonation sites on the molecule changes (see, e.g., \cite{de2016acidity}).

\item We know that the solute molecules in a solution must all aggregate in one place in order for a crystal to form.  To investigate the ease with which a crystal forms from a particular solution and the earliest stages of this nucleation process we might, therefore, calculate the free energy as a
function of the number of molecules that are in the solid phase
(see, e.g., \cite{cheng2015solid,tribello2017clustering}).
\end{itemize}

In all these examples the Hamiltonian depends on numerous degrees of freedom, but we are only interested in the behavior of a small number of these degrees of freedom (e.g., a torsional angle or a distance between two atoms).  A CV, $s$, is thus simply an arbitrary function of the atomic coordinates, $q$.
As we will see in what follows, in many cases the function $s$ is very simple and thus easy to calculate from $q$.  In other cases, however, the function is considerably more complicated so for these cases the scripting language of the PLUMED input file is invaluable. 

\subsection{Ensemble averages}
\label{sec:ensemble}

Section IV discussed a variety of different ways in which MD trajectories can be analyzed.
The simplest way to analyze the values a CV takes during a trajectory is to compute an ensemble average.  We are able to compute such averages from MD simulations because, if we are running at temperature $T$ on a system containing $N$ atoms that interact through a Hamiltonian, $H(\vec{q},\vec{p})$, the probability, $P(\vec{q},\vec{p})$, that a microstate in which the atoms have positions $\vec{q}$ and momenta $\vec{p}$ will be sampled at a particular instant in time is given by:
\begin{equation}
\label{eq:boltzmann}
P(\vec{q},\vec{p}) = \frac{ \exp\left( - \frac{H(\vec{q},\vec{p}) }{k_B T} \right) }{ \int \textrm{d}\vec{q}' \textrm{d}\vec{p}' \exp\left( - \frac{H(\vec{q}',\vec{p}') }{k_B T} \right) }
\end{equation}
where $k_B$ is Boltzmann's constant and where the $6N$-dimensional integral in the denominator runs over all the possible values of position and momentum that each of the atoms in the system might have.  Consequently, an observable $A(\vec{q})$ will have an ensemble average that is equal to:
\begin{equation}
\langle A \rangle = \int \textrm{d}\vec{q} \textrm{d}\vec{p} A(\vec{q}) P( \vec{q},\vec{p} )
\end{equation}
By the ergodic theorem, however, we know that if we add together the value $A(\vec{q})$ took in each of the frames in our molecular dynamics trajectory and if we divide this sum by the number of frames in our trajectory we  will obtain an estimate for $\langle A \rangle$.

\subsection{Free-energy landscapes}

A slightly more advanced way to analyze CVs is to compute the probability density along the CV. This function is defined as:

\begin{equation}
P(s)\propto\int d\vec{q}d\vec{p}P(\vec{q},\vec{p})\delta(s(\vec{q},\vec{p})-s)
\end{equation}
The free-energy surface, $F(s)$, is then nothing more than the negative logarithm of this probability density function expressed in units of energy:
\begin{equation}
F(s) = -k_B T \ln \left[ \int d\vec{q}d\vec{p}P(\vec{q},\vec{p})\delta(s(\vec{q},\vec{p})-s) \right] + C
\end{equation}
where $C$ is an arbitrary constant. 

\begin{figure}
\centering
\includegraphics[width=0.6\textwidth]{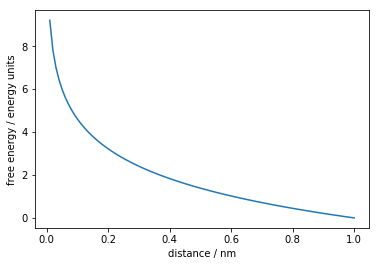}
\caption{Figure showing the free-energy surface as a function of the distance between two atoms that do not interact.  As you can see the free energy decreases as the two atoms move apart but this is only because the phase space volume that is accessible to two atoms that are exactly a distance $r$ apart is proportional to $r^2$.}
\label{fig:non-interacting}
\end{figure}

Recognizing this connection between the value of the thermodynamic potential and the probability of having a particular value for a collective variable is fundamental in terms of understanding what a free-energy landscape means.  As a case in point consider the free-energy surface shown in figure~\ref{fig:non-interacting}.  This figure shows how the free energy changes as the distance between two particles is increased.  One might be tempted, based on the shape of the curve, to assume that the two particles repel each other.  In actual fact, however, there is no interaction between the particles.  The free energy decreases because all possible distance vectors are equally likely. In three dimensions the probability of observing a particular distance, $r$, is thus proportional to $r^2$.  The free energy for a pair of non-interacting particles as a function of the distance between them is thus $F(r) = -2k_B T \ln r + C$.  Hence, when free-energy landscapes are used to provide information on the strength of the interaction between two particles, for example in a binding affinity study these two particles would be a small drug molecule and a protein, the surface obtained is usually corrected by adding
$2k_B T \ln r$ so that the free-energy surface for a pair of non-interacting particles appears flat.

\begin{figure}
\centering
\includegraphics[width=0.7\textwidth]{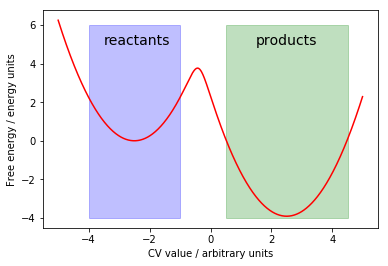}
\caption{Figure showing a typical one-dimensional free-energy surface that we might extract from an MD simulation.  It is clear from this diagram that there are two metastable basins in this landscape and that the system must cross a substantial barrier in order to get from one to the other.  The two shaded regions in this figure indicate the two sets of limits that we would integrate over when evaluating the free-energy difference between these two metastable basins using equation~\ref{eqn:fes-diff}.  To be clear, however, the precise choice for these limits will not affect the calculated free energy difference significantly as long as the two free-energy minima are reasonably deep.}
\label{fig:two-basins}
\end{figure}

Let us now suppose that we have extracted a free-energy surface that looks something like the one shown in figure \ref{fig:two-basins}.  This free-energy surface contains two metastable basins, which we will, for the time being, assume correspond to the reactant and product states for a chemical reaction that our simulated system can undergo.  If we want to calculate the free-energy change associated with this reaction we would use the expression below:  
\begin{equation}
F_{\textrm{products}}-F_{\textrm{reactants}}=-k_{B}T\log\left[ \frac{\int_{\textrm{products}}e^{-\frac{F(s)}{k_{B}T}}\textrm{d}s } {\int_{\textrm{reactants}}e^{-\frac{F(s)}{k_{B}T}}\textrm{d}s}  \right]
\label{eqn:fes-diff}
\end{equation}
Here the integrals in the numerator and the denominator of the quotient on the right-hand side run over the regions highlighted in figure \ref{fig:two-basins}.  Notice, furthermore, that the free-energy difference between the reactant and product states is calculated in this way because the value of the CV will fluctuate when it is in either state.  Calculating the reaction free energy using the formula above incorporates the effect of these fluctuations whereas simply calculating the difference between the values of the free energy at the bottom of two minima does not.

\begin{figure}
\centering
\includegraphics[width=0.7\textwidth]{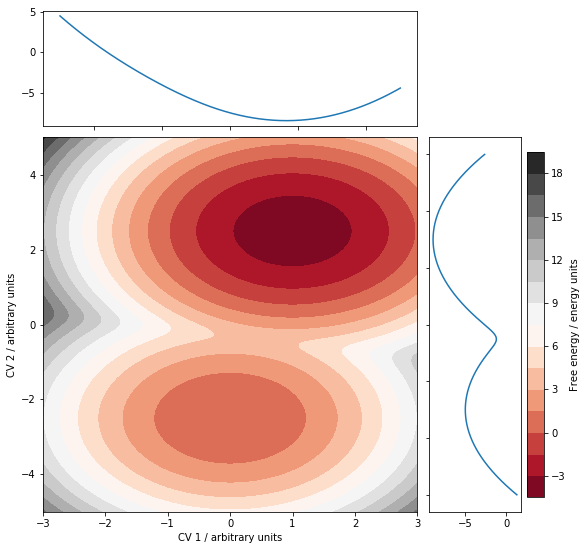}
\caption{The central panel of this figure shows a two-dimensional free-energy surface with two metastable states that is a function of two CVs called CV1 and CV2.  The panels above and to the right of this figure show the free-energy surface as a function of CV1 and CV2 respectively.  It is clear from these two figures that while CV2 is able to distinguish the two metastable basins CV1 is not.}
\label{fig:two-dimensional-surface}
\end{figure}

A further interesting thing that is worth noting about free-energy surfaces is that, if one has the free energy, $F(s)$, as a function of some CV, $s$, and if one wishes to determine the free energy as a function of some second CV, $\zeta$, as long as $\zeta$ is a one-to-one function of $s$ with an inverse function that is differentiable one can write an analytic expression for $F(\zeta)$ in terms of $F(s)$.  The reason that this is possible is that if $\zeta(s)$ is one-to-one then we know that:
\begin{equation}
P(\zeta) \textrm{d}\zeta = P(s) \textrm{d}s = P(s) \left | \frac{\textrm{d}s}{\textrm{d}\zeta}\right| \textrm{d}\zeta 
\end{equation}
Once we recall the connection between free energy and probability we thus arrive at:
\begin{equation}
F(\zeta)=F(s)-k_BT \ln \left | \frac{\textrm{d}s}{\textrm{d}\zeta}\right| 
\end{equation}
$F(\zeta)$ may look rather different to $F(s)$ but these two free-energy surfaces will still convey the same information about depths of the basins and the heights of the barriers.  In general, however, one should be particularly careful when discussing the barriers found in free-energy landscapes, as the height of the barrier will depend on the particular combination of CVs that are used to display the free-energy landscape.  As illustrated in figure \ref{fig:two-dimensional-surface}, if the critical degree of freedom that distinguishes between two metastable states is in a direction that is orthogonal to the CVs then these two states will both contribute their probability density to a single basin in the final free-energy landscape that is extracted.
If you want to extract information on  barriers you should thus probably
also extract the average passage times between states as this better characterizes the behavior of the chemical system.
Strictly speaking, a free-energy landscape alone can only tell you about the relative stabilities of the various metastable states that can be distinguished by the CVs that were used in its construction. Relating
the free-energy barriers to the rates with which the system will
pass from state to state is far from trivial and the result might depend
systematically on the capability of the CVs to correctly describe
the transition \cite{peters2016reaction}.

\subsection{Analyzing simulations with PLUMED}
\label{sec:examples}

In this section we will discuss how to compute various CVs using PLUMED.
It is important to emphasize at the outset that the PLUMED input files we provide can be used when running a simulation using some other MD code combined with PLUMED or when analyzing a simulation trajectory using PLUMED \texttt{driver}. In the first case the command line input will depend on which precise MD code one is using.
For instance, simulations done with GROMACS are typically launched using a command such as:
\begin{verbatim}
> gmx mdrun -plumed plumed.dat
\end{verbatim}
where \texttt{plumed.dat} is the name of the PLUMED input file. In the second case, however, as only PLUMED is being used, the command is nearly always the same and will be something akin to: 
\begin{verbatim}
> plumed driver --plumed plumed.dat --igro traj.gro
\end{verbatim}
where \texttt{traj.gro} is a trajectory file in GROMACS format and \texttt{plumed.dat} is once again a PLUMED input file.

For the final end user a PLUMED input file looks like it is written in a rudimentary and easy-to-use scripting language.  Each line in the input file tells the code to do something, which may be as simple as calculating a position of a center of mass or as complex as calculating and accumulating a metadynamics bias.  As these commands can be used in a wide range of different contexts and orders the user has the flexibility to do the full range of analyses described in the previous sections.  Furthermore, if they choose to incorporate some new functionality they can quickly start to use it in tandem with all of the other methods developed by members of the PLUMED community.

\subsection{Distances, angles and torsions}

\begin{figure}
\centering
\includegraphics[width=0.6\textwidth]{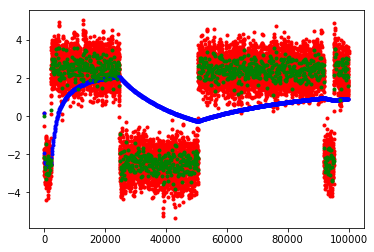}
\caption{Figure showing some things PLUMED can do with the CVs it calculates.  As discussed in the main text to calculate the curves shown in this figure we calculated the distance between atoms 1 and 2 on every 10th MD step.  The red points are thus the values of these distances.  The blue points indicate the mean value of this distance calculated for a range of differently sized samples.  Lastly, the green points are running averages from the first 100 frames, the second 100 frames and so on.}
\label{fig:d_vs_t}
\end{figure}

Chemical reactions are one class of phenomena that we can investigate using MD simulations.  In many chemical reactions two atoms or groups of atoms approach each other so that a chemical bond can form between them.  The ideal CV to use to describe this phenomenon is thus the distance between the relevant atoms.  We can calculate and print the distance between a pair of atoms using the PLUMED input file below:
\begin{verbatim}
d1: DISTANCE ATOMS=1,2
PRINT ARG=d1 FILE=colvar STRIDE=10
\end{verbatim}
This input instructs PLUMED to calculate the distance between the positions of the first and second atoms in the MD code's input file and to print this quantity to a file called \texttt{colvar} every 10 steps.  We can take the output from this calculation and plot a graph showing the value of the distance as a function of time.  The resulting curve would look something like the red curve in figure~\ref{fig:d_vs_t}. 

We can also use PLUMED to calculate the average value of the distance between two atoms.  The following input calculates the distance every 10 steps and then calculates a sample mean over the whole trajectory. %
\begin{verbatim}
d1: DISTANCE ATOMS=1,2
a1: AVERAGE ARG=d1 
PRINT ARG=a1 FILE=colvar STRIDE=100
\end{verbatim}
The output generated by this calculation is plotted in blue in figure~\ref{fig:d_vs_t}.  There are multiple points in the output file as the above input instructs PLUMED to calculate an ensemble average from the first 100 trajectory frames, the first 200 trajectory frames, the first 300 trajectory frames and so on.  It is worth noting that one can also use PLUMED to calculate the running averages from the first 100 frames, the second 100 frames and so on - points that are shown using green dots in figure~\ref{fig:d_vs_t}  - by using an input file like the one below:
\begin{verbatim}
d1: DISTANCE ATOMS=1,2
a1: AVERAGE ARG=d1 CLEAR=100
PRINT ARG=a1 FILE=colvar STRIDE=100
\end{verbatim} 

To calculate and print an angle between two bonds using PLUMED one might use an input file like the one below:
\begin{verbatim}
ang: ANGLE ATOMS=1,2,3 
PRINT ARG=ang FILE=colvar STRIDE=5
\end{verbatim}
As shown in figure \ref{fig:angles}(a) this input calculates the angle between the vector connecting atom 2 to atom 1 and the vector connecting atom 2 to atom 3.  The assumption here is that there are chemical bonds connecting atom 2 to atoms 1 and 3.  This input is thus measuring the angle between these bonds.  As illustrated in figure \ref{fig:angles}(b) we can, however, also use PLUMED to calculate the angle between any pair of distance vectors.  For example the input file below would calculate the angle between the vector that connects atom 2 to atom 1 and the vector that connects atom 3 to atom 4.
\begin{verbatim}
ang2: ANGLE ATOMS=1,2,3,4 
PRINT ARG=ang2 FILE=colvar STRIDE=2
\end{verbatim}

PLUMED can also be used to calculate and print the torsional angles illustrated in figure \ref{fig:angles}(c) that involve sets of four atoms as shown below:
\begin{verbatim}
tor: TORSION ATOMS=1,2,3,4
PRINT ARG=tor FILE=colvar STRIDE=1
\end{verbatim}

\begin{figure}
\centering
\includegraphics[width=0.9\textwidth]{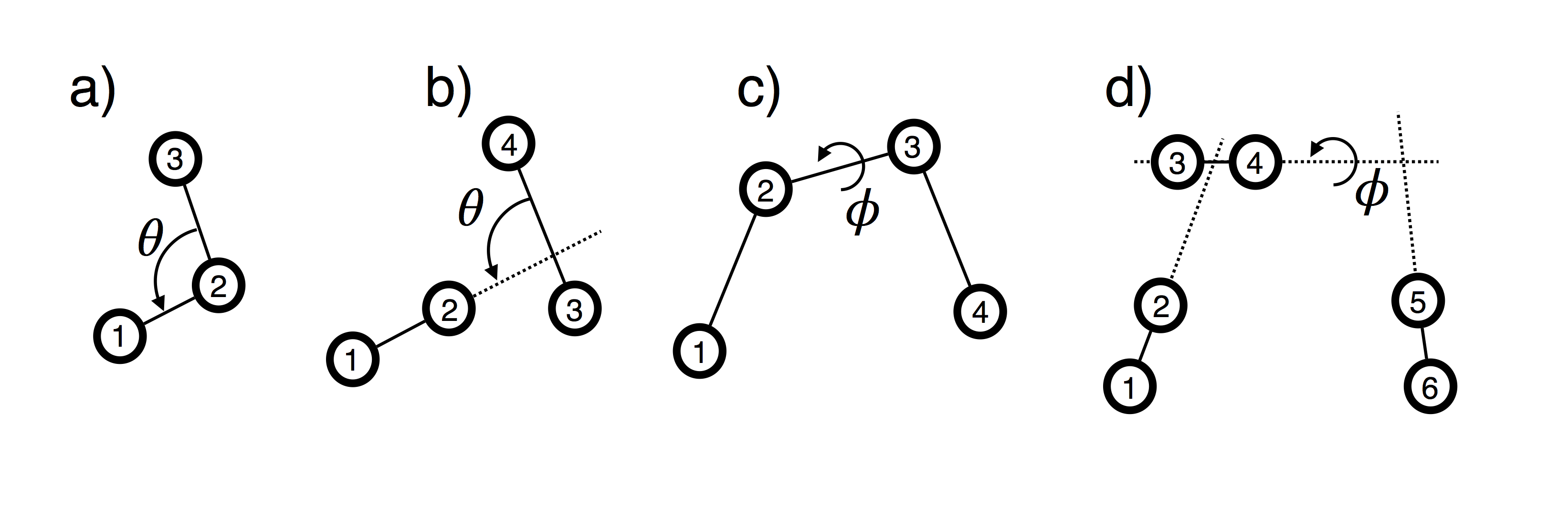}
\caption{Schematic representations of the angles and torsions that PLUMED can compute.
a) Angle defined using three atoms; b) Angle defined using four atoms; c) Torsion defined using four atoms;
d) Torsion defined using six atoms.}
\label{fig:angles}
\end{figure}

Much like the command to calculate the angle where three atoms are specified the assumption made when writing this input file is that there are chemical bonds between atoms 1 and 2, atoms 2 and 3 and atoms 3 and 4.  In general, however, a torsional angle measures the angle between two planes, which have at least one vector in common as illustrated in figure \ref{fig:angles}(d).  As shown below, there is thus an alternate, more general, way through which we can define a torsional angle:
\begin{verbatim}
tor2: TORSION VECTOR1=1,2 AXIS=3,4 VECTOR2=5,6
PRINT ARG=tor2 FILE=colvar STRIDE=20
\end{verbatim}
This input instructs PLUMED to calculate the angle between the plane containing the vector connecting atoms 1 and 2 and the vector connecting atoms 3 and 4 and the plane containing this second vector and the vector connecting atoms 5 and 6.  Notice that there is one additional syntax for selecting the atoms that should be used to calculate a torsion that is discussed in Note \ref{note-molinfo}.

\subsection{Positions and RMSD}

\begin{figure}
\centering
\includegraphics[width=0.9\textwidth]{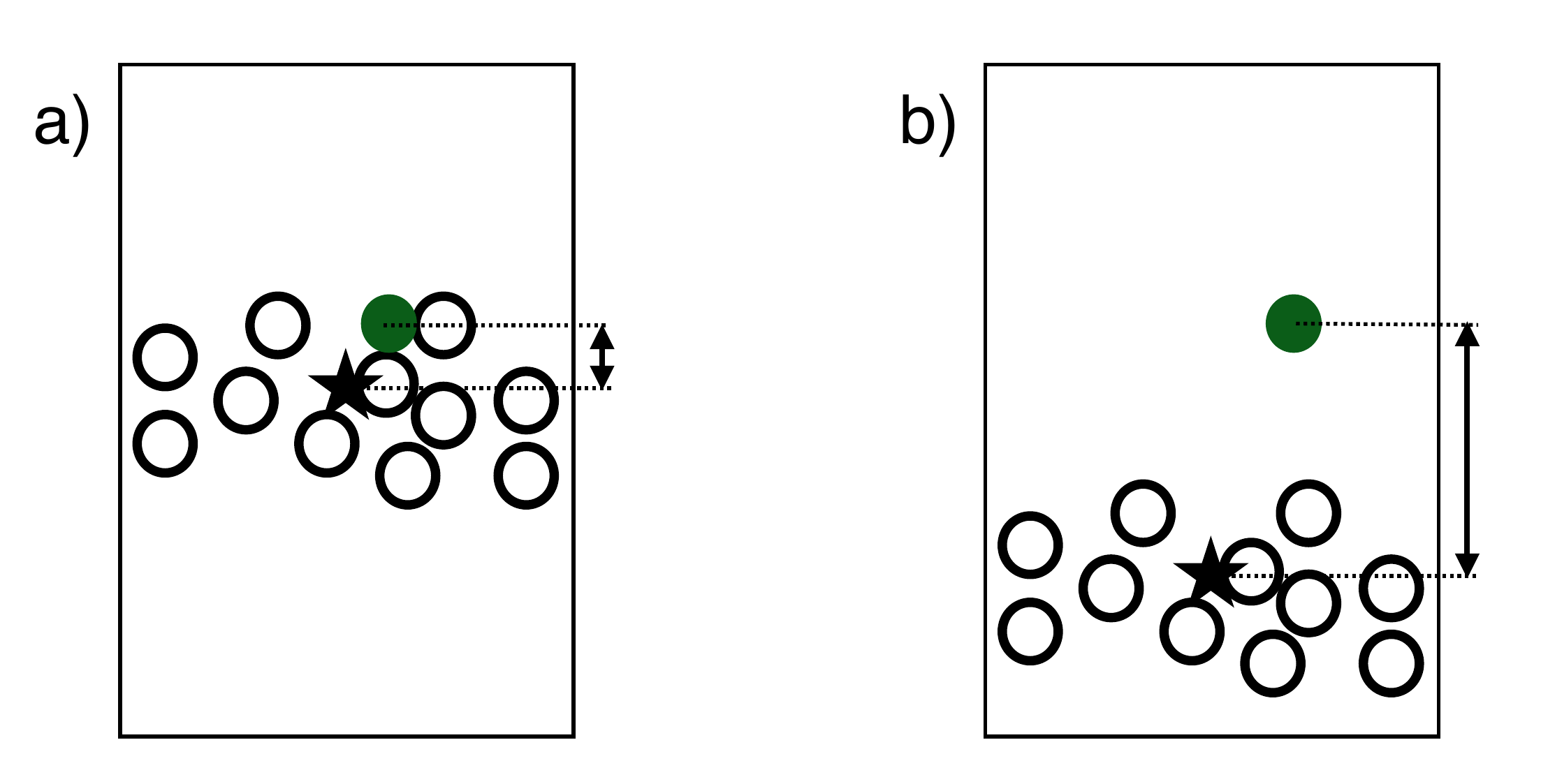}
\caption{Figure illustrating the vertical position of an ion with respect to a membrane.
The ion is shown in green, whereas the atoms of the membrane are depicted as empty circles.
The center of the membrane is represented using a star. Panels a) and b) report two different structures.
Notice that the vertical coordinate of the ion is the same, but that its position relative to the membrane
is different. The correct way to describe the position of the ion with respect to the membrane is thus to use
the vertical component of the distance between the ion and the center of the membrane.
}
\label{fig:ion-membrane}
\end{figure}
One of the processes that we stated we might want to study was the diffusion of ions across cell membranes.  
If the membrane is parallel to the $xy$ plane, one might be tempted to conclude that that correct CV to use in this case is the $z$ component of the position of the ion. Assuming that the ion is atom number 1001, this can be done with the following input file
\begin{verbatim}
pos: POSITION ATOM=1001
PRINT ARG=pos.z FILE=colvar STRIDE=10
\end{verbatim}
The variable \texttt{POSITION} that is used here has multiple components. As we want the \texttt{z} component specifically, we thus use \texttt{pos.z} in the \texttt{PRINT} command.  Using this CV is a bad idea as the potential energy functions that we use within MD simulations are almost always translationally invariant.  As illustrated in figure \ref{fig:ion-membrane} for our membrane example a particular value for the $z$ component of our ion's coordinate might be inside the membrane at one particular time. During the simulation, however, the membrane can move and as such this same value for the ion's $z$ coordinate might be outside the membrane at some later time. It is thus usually much better to use the $z$ component of the distance between the coordinates of the ion and the center of the membrane and to thus track the $z$ position of the ion relative to the $z$ position of the membrane.  An example input that can be used to calculate and print this quantity is given below:
\begin{verbatim}
c1: COM ATOMS=1-1000
dist: DISTANCE ATOMS=c1,1001 COMPONENTS
PRINT ARG=dist.z FILE=colvar STRIDE=10
\end{verbatim}
Here we assumed that the first 1000 atoms together form the membrane.

It is worth noting here how the COM command is used to keep track of the position of the center of mass of the large number of atoms that make up the membrane and how the position of this center of mass is referred to in the DISTANCE command using the label, c1, of the COM command.  It is also important to understand how PLUMED deals with periodic boundary conditions (see Section~\ref{sec:pbc}) and to remember that, for the position of the center to be computed correctly, the
vertical span of the membrane must be less than half of the box height. Finally, one should be aware that using a large number of atoms when calculating a CV may well slow down the calculation. It may thus be worth using only a subset of the atoms in the membrane when calculating the position of the center by, for instance, replacing the first line of the input file above with:
\begin{verbatim}
c1: COM ATOMS=1-1000:10
\end{verbatim}
The extra \texttt{:10} here tells PLUMED to only use every 10th atom in the range specified.  This is a crude solution, however, and it is probably always better in practice to select
specific atoms using your understanding of the chemistry of the system.

It can be useful to consider the atomic positions directly when considering problems such as protein folding.
Let's suppose that you know the precise arrangement that the atoms have in the native structure and that you want
to monitor the progression of folding during a trajectory.  An obvious CV to measure would be the degree of similarity between the instantaneous coordinates of the protein and this special, folded configuration.  A method that is commonly used to calculate this degree of similarity involves computing the root-mean-square deviation (RMSD) between the instantaneous coordinates and the coordinates in the reference structure using:
\begin{equation}
s=\sqrt{\frac{1}{N}\sum_{i=1}^N\left(\vec{q}_{i}-\vec{q}_{i}^{(\textrm{ref})}\right)^{2}}
\label{eqn:rmsd}
\end{equation}
In this expression $N$ is the number of atoms, $\vec{q}_{i}$ is used to denote the instantaneous coordinates of atom $i$ and $\vec{q}_{i}^{(\textrm{ref})}$ is used to denote the coordinates of atom $i$ in the reference structure. It is important to note that if you were to use the formula above you would have the same problem as if you used the position of an atom as a CV.  Consequently, the RMSD formula above is usually computed from a reference frame that is found by performing translation and rotation operations on the original reference structure that minimize the value of the RMSD \cite{kabsch1976solution}.
An input file that can be used to calculate and print this quantity using PLUMED is shown below:
\begin{verbatim}
rmsd: RMSD REFERENCE=ref.pdb TYPE=OPTIMAL
PRINT ARG=rmsd FILE=colvar STRIDE=5
\end{verbatim}
This input computes the root-mean-square deviation between the instantaneous positions of the atoms that are listed in the input ref.pdb file and the positions of those atoms in the pdb file.  In this case any translation of the center of mass and rotation of the reference frame is removed before calculating the displacements that enter equation \ref{eqn:rmsd}.  As discussed in Note \ref{note-rmsd}, however, PLUMED provides a range of options that allow you to perform these RMSD calculations in various different ways.

As we will see, PLUMED is often used to apply a simulation bias that is a function of the CVs it computes. The forces due to this bias are then propagated onto the atoms that were used to calculate the CV. 
When these CVs are calculated using the RMSD procedure outlined above the forces required to restore invariance with respect to translations and/or rotations
must be applied to all the atoms employed in the alignment procedure, which is computationally expensive.  In fact even if you don't have any forces to propagate just performing the alignment operation with a large number of atoms is expensive.
One would, therefore, typically never use the positions of all the atoms when performing RMSD calculations.

An alternative to RMSD is the so-called DRMSD:
\begin{equation}
s = \sqrt{ \frac{1}{M}\sum_{ij} \left( d_{ij} - d_{ij}^{(\textrm{ref})} \right)^2 } 
\label{eqn:drmsd}
\end{equation}
In this expression the sum runs over $M$ pairs of atoms and $d_{ij}$ is used to denote the instantaneous distance between these pairs while $d_{ij}^{(\textrm{ref})}$ is used to denote the distance between the corresponding pair of atoms in the reference structure.  An input file that calculates and prints this quantity using PLUMED is given below.  Notice how the two cutoff keywords can be used to specify the range of values the distances should take in the reference structure in order to be considered as part of the sum in equation \ref{eqn:drmsd}.
\begin{verbatim}
# the dots here indicate that the command
# will be continued on the following line.
drmsd: DRMSD ...
  REFERENCE=ref.pdb LOWER_CUTOFF=0.1 UPPER_CUTOFF=0.8
...
PRINT ARG=drmsd FILE=colvar STRIDE=5
\end{verbatim}

\subsection{Gyration radius and gyration tensor}

The CVs that have been discussed thus far are all based on a very detailed view of the positions of particular atoms.  CVs that give a more coarse-grained view of the instantaneous shape of a molecule can also be used, however.  One particularly popular CV of this type is the so-called radius of gyration, which can be calculated using:
\begin{equation}
s=\sqrt{\frac{1}{\sum_i w_i }\sum_{i} w_i \left(\vec{q}^{(i)}-\vec{q}^{(c)}\right)^{2}}
\label{eqn:gyr}
\end{equation}
where $\vec{q}^{(c)} = \frac{\sum_i w_i \vec{q}^{(i)} }{ \sum_i w_i }$ and $\vec{q}^{(i)}$ is the position of the $i$th atom.  The $w_i$ are a set of weights that are ascribed to each of the atoms in the system.  These weights might be, for instance, the masses of the atoms.  To calculate and print this quantity using PLUMED you would use an input like the one below:
\begin{verbatim}
gyr: GYRATION TYPE=RADIUS ATOMS=10-20 
PRINT ARG=gyr STRIDE=1 FILE=colvar
\end{verbatim}
This calculates the radius of gyration using the positions of the 10th to the 20th atom in the MD code's input file and sets the weights of all these atoms equal to 1. 

\begin{figure}
\centering
\includegraphics[width=0.9\textwidth]{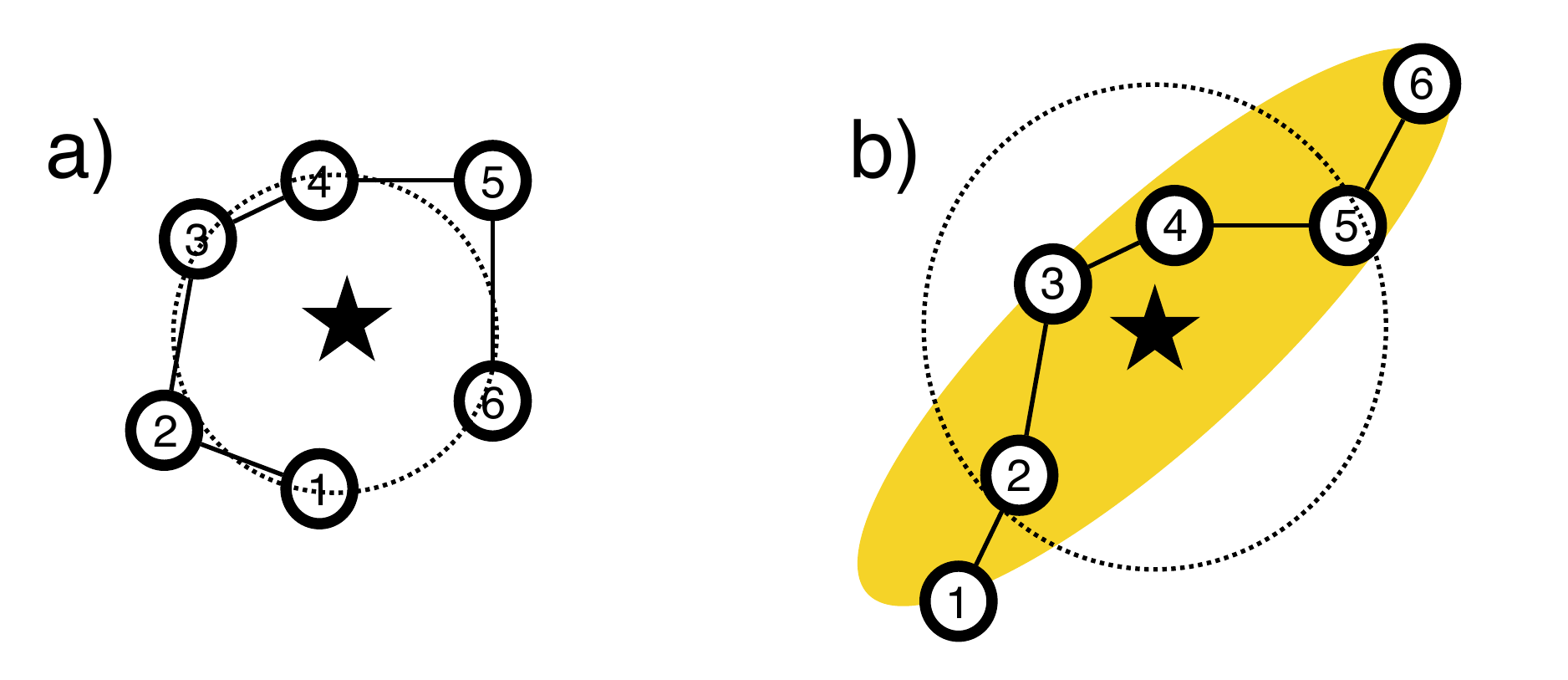}
\caption{Figure illustrating the components of the radius of gyration of a polymer.
The center of the polymer is represented as a star. A circle with a radius corresponding to the radius
of gyration of the polymer and centered on the center of the polymer is shown using a dashed line.
a) For a globular polymer, the circle correctly represents the shape of the polymer.
b) For an elongated polymer, however, the circle does not represent the shape of the polymer correctly.
An ellipsoid with axes that have lengths corresponding to the square roots of the eigenvalues of the gyration tensor
is more representative and is shown in yellow.
}
\label{fig:polymer}
\end{figure}

When the gyration radius is calculated using equation~\ref{eqn:gyr} the quantity output provides a measure of the average radius of the molecule.  No information on the shape of the molecule is provided, however, and so this average radius may be misleading.  For example, and as shown in figure~\ref{fig:polymer}, the radius of gyration for an extended polymer, which has a shape that is very anisotropic, would not be representative of the extent of the molecule in either its extended or compact directions.  For this reason, some researchers have chosen to use the eigenvalues of the gyration tensor instead \cite{vymetal2011gyration}.  The elements of the $3 \times 3$ gyration tensor are computed using:
\begin{equation}
s_{jk} = \frac{1}{\sum_i w_i } \sum_{i} w_i \left(q^{(i)}_j - q^{(c)}_j \right) \left(q^{(i)}_k - q^{(c)}_k \right) 
\end{equation}
where $q^{(i)}_j$ is used to denote the $j$th component of the position of atom $i$ and $q^{(c)}_j$ is used to denote the $j$th component of $\vec{q}^{(c)}$.  Figure~\ref{fig:polymer} shows how the square roots of the eigenvalues of this matrix give a sense of the shape of the molecule.  By changing the word after the TYPE keyword in the PLUMED input above one can access these eigenvalues directly or various functions of these eigenvalues that can be used to give one a sense of the sphericity of the molecule or the cylindricity \cite{vymetal2011gyration}.

\begin{figure}
\centering
\includegraphics[width=0.9\textwidth]{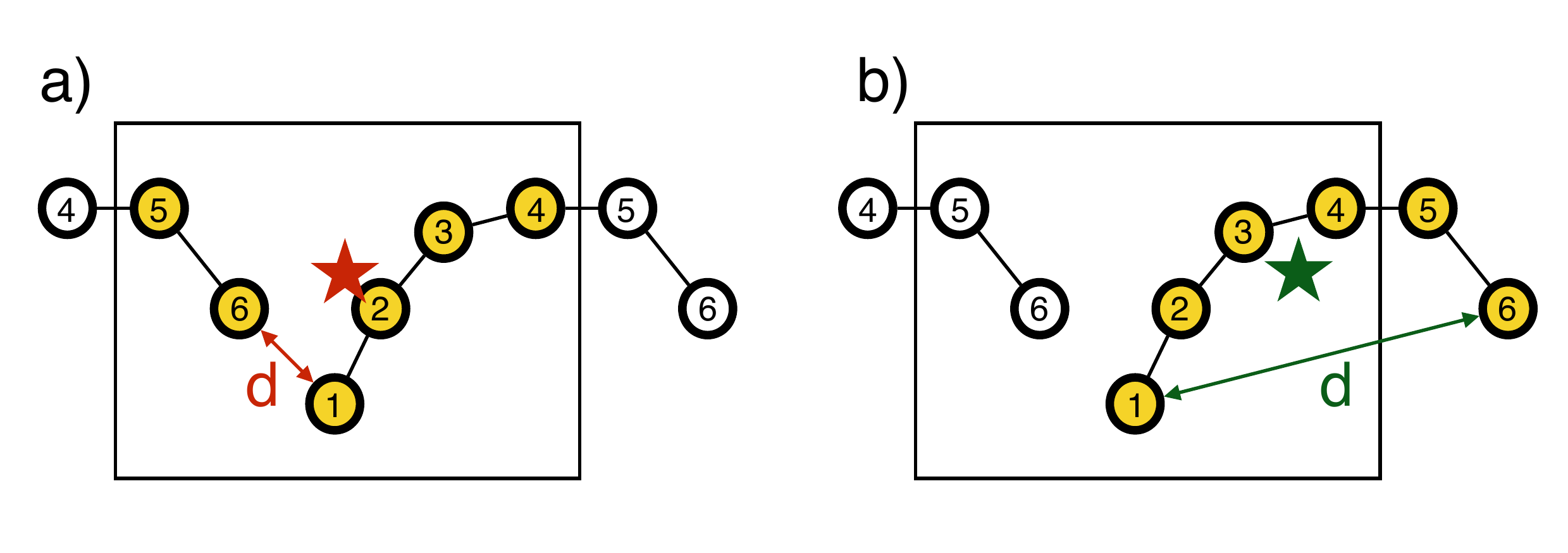}
\caption{
An illustration showing how molecules can be split by the periodic boundary conditions and how this can cause a problem when computing
collective variables.  Panels a) and b) represent the same set of six atoms.  In both of these figures these atoms sit in the periodic box indicated using the rectangle so  periodic images of atoms 4, 5, and 6 are included. The atoms used to calculate the collective
variables are highlighted. In panel a), a broken molecule is used so the position of center of mass (indicated using a star) and the end-to-end distance (indicated using an arrow) are computed incorrectly. In panel b), however, the molecule has been
correctly reconstructed across the periodic boundaries using \texttt{WHOLEMOLECULES} and the center of mass has thus been computed correctly. Notice that the correct value for the end-to-end distance is only obtained from panel (b) if the periodic boundary conditions are \emph{ignored} when computing the distance (keyword \texttt{NOPBC}).  If the PBC are taken into account the incorrect image of atom 6 will be used and the incorrect result illustrated in  panel (a) will be obtained once more.
}
\label{fig:wholemolecules}
\end{figure}

\subsection{Dealing with periodic boundary conditions}

\label{sec:pbc}
Figure \ref{fig:wholemolecules} illustrates a technical problem that can appear when PLUMED is used to calculate some CVs.  When the underlying MD code applies the periodic boundary conditions molecules can end up split across either side of the periodic box.  Consequentially, atoms can appear to be much farther apart than they are in actuality and, as shown in the figure, the position of the center of mass of the molecule can be calculated wrongly.  In this case, however, and in some of the others discussed thus far PLUMED resolves this problem automatically by adjusting the positions so that the set of molecules that are used to calculate the position of a center of mass or a CV form one single unbroken molecule.  This is still a problem that any user must be aware of, however, as there are some cases that PLUMED cannot fix automatically.  For example suppose that you wanted to calculate the end to end distance for the molecule illustrated in \ref{fig:wholemolecules}.  In order to do this correctly one must reconstruct the whole molecule before calculating the distance between the two terminal atoms.  To resolve this problem PLUMED provides a command, WHOLEMOLECULES, that allows one to adjust the way the positions are stored and to thus specify the molecules that must be reconstructed.  A sample input that calculates this end to end distance and that uses a WHOLEMOLECULES command is provided below
\begin{verbatim}
WHOLEMOLECULES ENTITY0=1-6
d1: DISTANCE ATOMS=1,6 NOPBC
\end{verbatim}
The \texttt{WHOLEMOLECULES} command here ensures that the bond between each pair of adjacent atoms specified to the \texttt{ENTITY} keyword is not broken by the periodic boundaries.  The distance is thus computed from the positions of the atoms shown in the right panel of figure~\ref{fig:wholemolecules}. There is thus no need to apply periodic boundary conditions.  In fact if, when the molecule is extended, it has a length that is longer than half the box length it is wrong to apply periodic boundary conditions as the ``end-to-end distance" computed this way would no longer be representative of the distance along the chain.   

\begin{figure}
\centering
\includegraphics[width=0.9\textwidth]{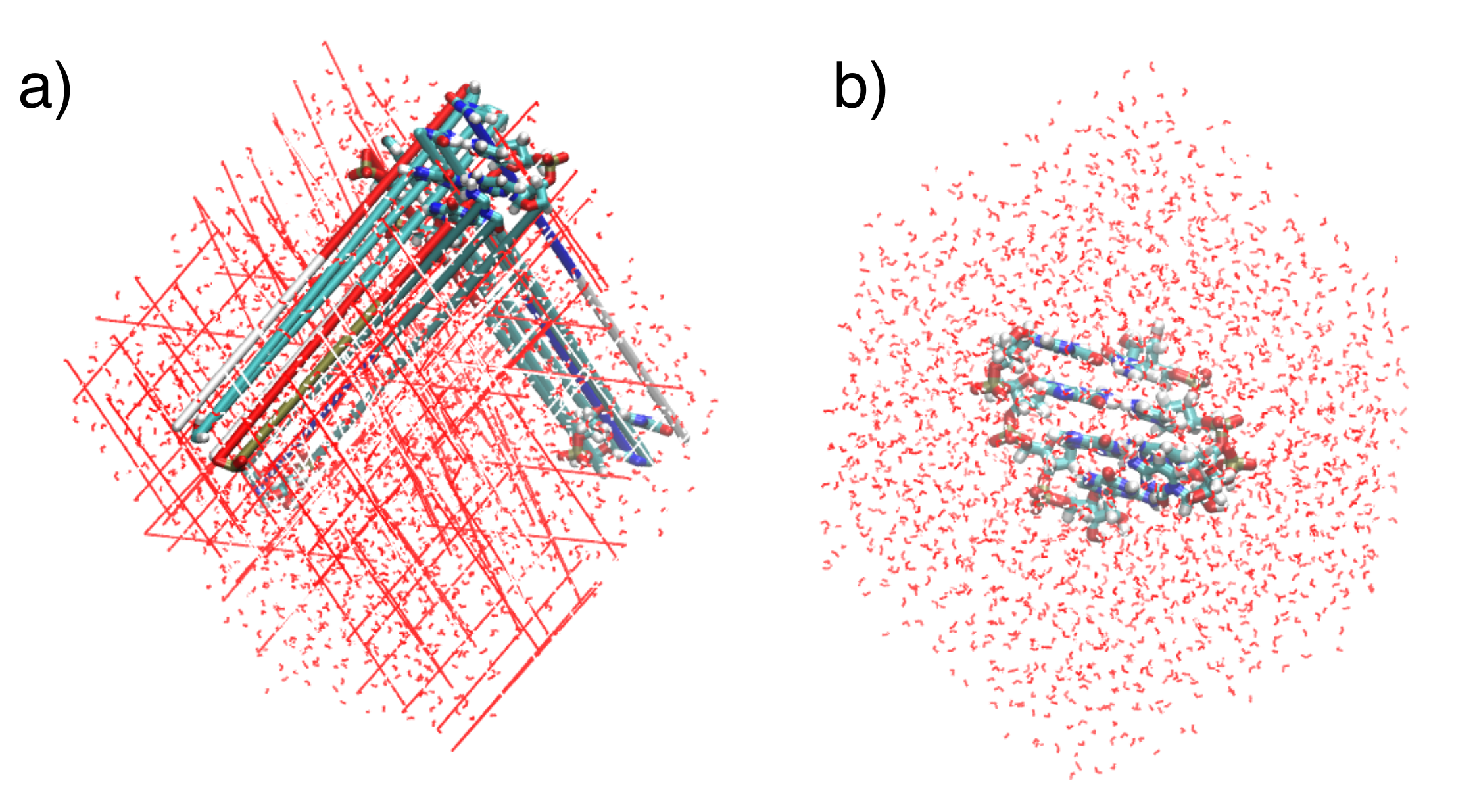}
\caption{
Figures showing how PLUMED can be used to reconstruct a solvated RNA molecule that has been split by the periodic boundary conditions. The RNA duplex is represented using licorice and the
water molecules are represented using lines. In panel (a) the molecules that cross the periodic boundaries are broken. In panel (b), however, the RNA molecule has been reconstructed and centered, and the periodic
images of the water molecules that are closest to the center of the RNA molecule have been selected. Notice that selecting the water molecules in this way ensures that the water molecules are made whole.
}
\label{fig:wholemolecules-complex}
\end{figure}

A more complicated example which has been taken from Ref.~\cite{cunha2017unraveling} and which also requires the \texttt{WHOLEMOLECULES} command to be used is shown in figure~\ref{fig:wholemolecules-complex}.
In this case the reconstruction has been done using the following input file
\begin{verbatim}
MOLINFO STRUCTURE=ref.pdb

rna: GROUP ATOMS=1-258
mg:  GROUP ATOMS=6580
wat: GROUP ATOMS=259-6579

# Make the RNA duplex whole.
WHOLEMOLECULES ENTITY0=rna

# Align the RNA duplex to a reference structure
FIT_TO_TEMPLATE REFERENCE=ref-rna.pdb TYPE=OPTIMAL
# Notice that before using FIT_TO_TEMPLATE
# we used WHOLEMOLECULES to make the RNA whole
# This is necessary otherwise you would be aligning
# to a broken molecule!

# compute the center of the RNA molecule
center: CENTER ATOMS=rna

# Wrap atoms correctly around the center of the RNA
WRAPAROUND ATOMS=mg AROUND=center
WRAPAROUND ATOMS=wat AROUND=center GROUPBY=3 
# Dump the resulting trajectory
DUMPATOMS ATOMS=rna,wat,mg FILE=rna-wrap.gro
\end{verbatim}

Notice here how the action \texttt{FIT\_TO\_TEMPLATE} has been used to align the RNA molecules to a template structure
that is at the center of the box and how the action \texttt{WRAPAROUND} has been used to reposition the water molecules around the aligned RNA molecule.  In addition, notice that by aligning the molecule to a template we have made the positions of these atoms roto-translationally invariant.  We can thus use the positions of the aligned atoms as CVs using the \texttt{POSITION} command that was introduced earlier. Furthermore, when we do so the \texttt{FIT\_TO\_TEMPLATE} command will ensure that the process of aligning the molecule is considered correctly when propagating any forces to the underlying positions.

As it is probably clear at this stage,  correctly reconstructing the atoms across the periodic boundary conditions is crucial as when this is not done some variables will be computed incorrectly. In order to simplify the preparation of PLUMED input and to decrease the number of errors, we have tried to automatize the reconstruction in all the cases where this is easy to do. You should thus check the manual of the PLUMED version that you are using to know when the reconstruction of molecules that have been broken by the periodic boundary conditions is dealt with automatically.

\subsection{Going further with collective variables}

Hopefully the example input files in the previous sections have given you a sense of how PLUMED input files work.  In essence, all the PLUMED commands for calculating CVs that we have introduced thus far calculate a scalar valued quantity.  Any scalar valued quantity that we calculate can then be referred to later in the input file by using the label of the command that calculates it.  So for instance in the following PLUMED input file:
\begin{verbatim}
dist: DISTANCE ATOMS=1,2
PRINT ARG=dist FILE=colvar STRIDE=1
\end{verbatim}
the \texttt{PRINT} command instructs PLUMED to print the quantity called dist which is calculated by the first \texttt{DISTANCE} command to a file called colvar during every MD step.  The fact that we can pass scalar valued quantities in this way is enormously useful as it means we can script new CVs directly from the input file.  For example suppose that we want to calculate the number of native contacts in a protein.  This CV is often computed using the continuous switching function shown below or by using some other function that displays a similar behavior \cite{best2013native}:
\begin{equation}
s = \sum_{ij \in N_C} \frac{1}{1 + (r_{ij}/r_0)^6}
\end{equation}
Here the sum runs over the list of pairs of atoms that are in contact when the protein is folded and $r_0$ is a parameter.  If we had a total of four native contacts in the protein and if $r_0$ were set equal to 6 \AA\ we could compute and print this quantity using the following PLUMED input file.
\begin{verbatim}
d1: DISTANCE ATOMS=1,2
d2: DISTANCE ATOMS=5,6
d3: DISTANCE ATOMS=9,10
d4: DISTANCE ATOMS=15,16
# The braces allow us to use spaces
# within the argument of the FUNC keyword
contacts: CUSTOM ...
  ARG=d1,d2,d3,d4 VAR=a,b,c,d
  FUNC={
    1/(1+(a/0.6))^6
   +1/(1+(b/0.6))^6
   +1/(1+(c/0.6))^6
   +1/(1+(d/0.6))^6
  }
  PERIODIC=NO
...
PRINT ARG=contacts FILE=colvar STRIDE=10
\end{verbatim}
Notice here how we have used our familiar friend the \texttt{DISTANCE} command to calculate each of the distances required and have then used the command \texttt{CUSTOM} to calculate the non-linear combination of quantities that the CV requires.  The fact is that most of the more complicated CVs that have been used to analyze molecular dynamics trajectories are simply linear or non-linear combinations of the quantities that have been introduced thus far.  One can thus compute many complicated CVs by just using the commands that were introduced in the previous sections together the with \texttt{CUSTOM} command.  It is worth understanding how to use this approach, but we would, in practice, not recommend you write your PLUMED input files in this way.  Instead, we would recommend that you use the numerous shortcuts PLUMED provides for accessing these non-linear combinations.  For example a shorter input that does the same calculation as the input file above is as follows:
\begin{verbatim}
dists: DISTANCES ...
  ATOMS1=1,2 ATOMS2=5,6 ATOMS3=9,10 ATOMS4=15,16
  LESS_THAN={RATIONAL R_0=0.6}
...
PRINT ARG=dists.lessthan FILE=colvar STRIDE=10
\end{verbatim}
This input takes advantage of the fact that many of the more complicated CVs that we wish to compute have a functional form that can be thought of as follows:
\begin{equation}
s = \sum_{i=1}^N f(\{X\}_i)
\end{equation}
In other words, to calculate the CV one computes the same function, $f$, for $N$ different sets, $\{X\}_i$, of atomic positions.  Some more complicated examples of CVs of this form are the so-called secondary structure variables \texttt{ALPHARMSD}, \texttt{ANTIBETARMSD} and \texttt{PARABETARMSD} \cite{pietrucci2009collective}.
For \texttt{ALPHARMSD} one takes each set of $M$ atoms in the protein that might together be able to form an alpha helix.  For each of these sets of variables one then computes the DRMSD distance between the instantaneous positions of the atoms and the positions of the atoms in an ideal alpha helix.  Each of these DRMSD distances is then transformed by a continuous switching function and these transformed values are then added together.  The final value of the CV, that is calculated and printed using the input below, thus measures how many segments in the protein resemble an alpha helix. 
\begin{verbatim}
MOLINFO STRUCTURE=helix.pdb
a: ALPHARMSD ...
  RESIDUES=all TYPE=DRMSD 
  LESS_THAN={RATIONAL R_0=0.08 NN=8 MM=12} 
...
PRINT ARG=a.lessthan FILE=colvar STRIDE=10
\end{verbatim}
The \texttt{ANTIBETARMSD} and \texttt{PARABETARMSD} commands do something very similar but the reference configuration in these cases are obviously an ideal anti-parallel beta sheet and an ideal parallel beta sheet respectively.  For these three commands one could, in theory, calculate this quantity using multiple 
\texttt{DRMSD} commands and the \texttt{CUSTOM} command that was introduced earlier. It is obviously much easier though to use the input above, which determines the sets of atoms for which you have to determine DRMSD distances from the template
pdb structure directly.  Having said that, however, it can be useful to try to write input files that use only the simple commands that were introduced in the previous sections in order to better understand what precisely is being calculated by these shortcuts.   

This chapter would be overly long if we listed all the CVs that are available in PLUMED.  If you are interested we would recommend reading the literature.  To get you started we have provided  
the following short, but far from exhaustive list, of references for a range of CVs, which can be calculated by PLUMED.  This list includes
the total energy of the system \cite{bartels1998probability,bonomi2010enhanced},
some of the components of the energy \cite{lazaridis1999effective,do2013rna},
the dimer interaction energy \cite{nava2017dimer},
discrepancy measures \cite{bottaro2016free},
principal components \cite{spiwok2007metadynamics,sutto2010comparing},
path collective variables \cite{branduardi2007b,leines2012path},
property maps \cite{spiwok2011metadynamics},
puckering variables \cite{cremer1975general,huang2014improvement},
Steinhard order parameters \cite{tribello2017clustering} and
a number of experimental observables \cite{bonomi2017integrative}.  In addition, PLUMED contains implementations of the principle component analysis \cite{PCA_book}, multidimensional scaling \cite{MDS_book} and sketch-map \cite{sketchmap} algorithms, which are all tools that can be used to analyze simulation trajectories.  PLUMED input files can be prepared
using a graphical user interface \cite{giorgino2014plumed} and this graphical user interface also allows you to compute PLUMED CVs from within
the VMD program \cite{humphrey1996vmd}.

\section{Biasing collective variables}
\label{sec:biasing}

Section \ref{sec:cvs} has hopefully given you a sense of some of the quantities we might be interested in monitoring during the course of a simulation.    If this were all that PLUMED could do, however, there would be no reason for it to be usable on the fly.  After all, all of the quantities we have discussed can be calculated by post-processing the trajectory files that are output during the simulation. The reason that PLUMED runs on the fly is that it can also modify the Hamiltonian, $H(\vec{q},\vec{p})$.  These modifications introduce additional forces that must be incorporated when the underlying MD code integrates the equation of motion.  Calculation of the PLUMED potential and the associated forces can, however, be separated from the calculation of the forces due to the underlying potential as the bias potential, $V(\vec{q},t)$, that is calculated by PLUMED and that is a function of the atomic positions and possibly also time, $t$, is simply added to the Hamiltonian, $H(\vec{q},\vec{p})$, that is calculated by the MD code.  In other words, the modified Hamiltonian, $H'(\vec{q},\vec{p})$ is:
\begin{equation}
H'(\vec{q},\vec{p}) = H(\vec{q},\vec{p}) + V(\vec{q},t)
\end{equation}

\subsection{Reweighting}
\label{sec:reweight}

To understand how free energies can be extracted from biased MD simulations we must return once again to the probability distribution that is sampled during a molecular dynamics trajectory.  This distribution was first introduced in section \ref{sec:ensemble} when we introduced the following equation: 
\begin{equation}
P(\vec{q},\vec{p})\propto e^{-\frac{H(\vec{q},\vec{p})}{k_B T} } 
\label{eqn:unbiased}
\end{equation}
and therefore explained how the quantity on the right-hand side of this equation is proportional to the probability of sampling any given vector of atomic positions, $\vec{q}$, and momenta, $\vec{p}$.
Now suppose that we are not integrating the Hamiltonian $H(\vec{q},\vec{p})$ given above and that we are instead integrating a modified Hamiltonian $H(\vec{q},\vec{p})+V(\vec{q})$.  The probability distribution, $P'(\vec{q},\vec{p})$, that we will sample from when we integrate this biased Hamiltonian will be proportional to:
\begin{equation}
P'(\vec{q},\vec{p})\propto e^{-\frac{H(\vec{q},\vec{p})}{k_B T} } e^{-\frac{V(\vec{q})}{k_B T} }
\label{eqn:biased-probability}
\end{equation}
for the same reasons that the probability of sampling a particular configuration when the system is unbiased is given by equation~\ref{eqn:unbiased}.  Notice, furthermore, that the right-hand side of equation~\ref{eqn:unbiased} appears in the equation above and that we can thus rewrite this expression as:
\begin{equation}
P(\vec{q},\vec{p}) \propto P'(\vec{q},\vec{p})e^{+\frac{V(\vec{q})}{k_B T} } 
\label{eqn:reweight}
\end{equation}
This equation relates the probabilities that we extract from biased simulations to the corresponding unbiased probabilities and, as we will see in the next section, it is thus the central equation for the analysis of biased simulations. Strictly speaking, this relationship is only valid
if the bias potential does not change with time.  As discussed below,  different formalisms can be used when the bias potentials are time dependent.

\subsection{Extracting the free energy}
\label{sec:histograms}

We now have all the pieces we require and can thus finally discuss the topic that was first introduced in section \ref{sec:cvs}; namely, how we analyze biased and unbiased MD simulations so as to extract the free energy as a function of a small number of collective variables.  The first step in this process is to discretise the CV space and to introduce a set of probabilities, $p_j$, each of which tells us the probability that the CV value falls in a particular range or bin.  These bins are set up so that all possible CV values fall within exactly one of the bins.  Consequently, if we discretize the CV space, $s(\vec{q_i})$, into $M$ bins
with centers at $s_j$ the probability that $t_1$ of the frames from our MD trajectory fall within the first of these bins, that $t_2$ of frames fall within the second bin and so on is given by the following multinomial distribution: 
\begin{equation}
P(t_1,t_2,\dots,t_M) \propto \prod_{j=1}^M p_j^{t_j}
\label{eqn:multinom}
\end{equation}
where $p_j$ is the probability that any given trajectory frame will fall into the $j$th bin.  It is easy to extract the set of $t_j$ values that appear in this expression from our trajectory.  All we need to do is construct a histogram that tells us the number of times the trajectory visited each bin (see Note \ref{note-histogram}).  Furthermore, once we have this information on the values of $t_j$ we can perform a constrained optimization on equation \ref{eqn:multinom} and thus determine the most likely values for the probabilities, the $p_j$s, given the particular set of $t_j$ values that we observed during our trajectory.  It is easier if we start this process of optimizing equation \ref{eqn:multinom} by taking the logarithm of both sides of the equation.  Doing so has no effect on the position of the minimum as the logarithm is a monotonically increasing function.
Having taken the logarithm we then recall that the set of $p_j$ values are probabilities and that as such they must satisfy $\sum_{j=1}^M p_j = 1$.  We must therefore perform a constrained optimization using the method of Lagrange multipliers.  The final function we need to optimize is thus:
\begin{equation}
\mathcal{L}
\label{eq:lagrangian1}
= \sum_{j=1}^M t_j \ln p_j + \lambda\left( \sum_{j=1}^M p_j - 1 \right)
\end{equation}
where $\lambda$ is a Lagrange multiplier.
When this expression is maximized we obtain (see Note \ref{note-lagrangian})
\begin{equation}
p_k = \frac{t_k}{T}
\end{equation}
where $T=\sum_j t_j$ is the total number of frames in the trajectory.
In other words, the most likely estimate for the probability that the CV will take a value within a particular range is just the fraction of time that the trajectory spent with CV values in that particular range.  By recalling the relationship between probability and free energy we can thus express the free energy for having the CV value in the $k$th range as:
\begin{equation}
F_k = - k_B T \ln t_k + C
\end{equation}

Let us now suppose that we had taken the data from a biased trajectory rather than an unbiased trajectory and discuss how we would extract the histogram in this case.  In other words, suppose that instead of integrating the Hamiltonian, $H(\vec{p},\vec{q})$, directly we had integrated the biased Hamiltonian, $H(\vec{p},\vec{q}) + V(\vec{q})$.  As discussed in the previous section we know that the probability distribution that we sample configurations from when we do this biased simulation, $P'(\vec{q},\vec{p})$, is related to the probability distribution, $P(\vec{q},\vec{p})$, that we would have sampled configurations from had we run our simulation without the bias  by equation~\ref{eqn:reweight}.  We therefore might expect that we can extract a free-energy landscape for the unbiased simulation using the data from our biased simulations and a procedure much like that outlined above.  The trick for doing this is to recognize that we can use equation \ref{eqn:reweight} to write the multinomial distribution that we sample from when we run a \emph{biased} simulations in terms of the elements of the probability distribution, $p_j$, that would have been sampled if we had run our simulation \emph{without any bias} as shown below:
\begin{equation}
P'(t_1,t_2,\dots,t_M) \propto \prod_{j=1}^M (w_j p_j)^{t_j},
\end{equation}
where $w_j=e^{-\frac{V(\vec{q}_i)}{k_BT}}$
and where $V(\vec{q}_i)$ is the bias potential calculated for the $i$th frame.
If we take the logarithm of this expression and impose the constraint that $\sum_{j=1}^M (w_j p_j) = 1$ we thus find that the function to optimize in this case is:
\begin{equation}
\label{eq:lagrangian2}
\mathcal{L}= \sum_{j=1}^M t_j \ln w_jp_j + \lambda \left( \sum_{j=1}^N w_j p_j - 1 \right)
\end{equation}
When this expression is maximized we obtain (see Note \ref{note-lagrangian-weight}):
\begin{equation}
p_k \propto t_k e^{+\frac{V(\vec{q_i})}{k_B T}}
\label{eqn:weighted-histogram}
\end{equation}
By taking the logarithm of the above and by multiplying by $-k_B T$ we can thus extract the unbiased free-energy surface from a biased simulation.

\subsection{Biasing simulations with PLUMED}

The need for numerous methods based on integrating modified Hamiltonians has been discussed at length in section II.  We will thus limit the discussion of these methods here to providing a few sample PLUMED input files that instruct PLUMED to calculate various bias potentials.  The first of these examples involves the following input, which instructs PLUMED to use a harmonic restraint on the distance between atoms 1 and 2 as a bias potential.  The instantaneous values for the distance and the bias potential are then output to a file called colvar.  
\begin{verbatim}
d1: DISTANCE ATOMS=1,2
rr: RESTRAINT ARG=d1 AT=0.2 KAPPA=10 
PRINT ARG=d1,rr.bias FILE=colvar 
\end{verbatim}
This input encourages the system to sample configurations where the distance between atoms 1 and 2 is close to 0.2~nm.  Many such inputs are typically used when performing umbrella sampling calculations with multiple restraints
\cite{torrie1977nonphysical,kumar1992weighted}.

Bias potentials do not have to be independent of time.  There are, in fact, a whole class of steered MD methods in which a time dependent bias potential is used to force the system to change its conformation over the course of a simulation.  A sample PLUMED input for such a calculation is given below.
\begin{verbatim}
MOLINFO MOLINFO STRUCTURE=helix.pdb
phi3: TORSION ATOMS=@phi-3
mr: MOVINGRESTRAINT ...
   ARG=phi3 STEP0=0 STEP1=10000 STEP2=20000
   KAPPA=100 AT0=-pi/3 AT1=pi/4 AT2=-pi/3
... 
PRINT ARG=phi3,mr.* FILE=colvar
\end{verbatim}
The input above is for a 20000 step steered MD simulation \cite{isralewitz1997binding} in which the $\phi$ angle in the third residue of protein is forced to change its value from $-\frac{\pi}{3}$ radians to $\frac{\pi}{4}$ radians before being forced to change the value of this torsional angle back to $-\frac{\pi}{3}$ radians once more.  This input instructs PLUMED to output the instantaneous value of the torsional angle, the bias and the work the bias has done on the system.  Obviously, the user can make the path the system is forced to take through configuration space more complicated by using linear combinations of CVs and by changing the number of STEP and AT commands.

The final sample input below gives an example of how PLUMED can be used to perform the metadynamics simulations \cite{laio2002escaping} that were discussed at length in Chapter 4.
\begin{verbatim}
phi: TORSION ATOMS=5,7,9,15
psi: TORSION ATOMS=7,9,15,17
metad: METAD ...
  ARG=phi,psi PACE=500 HEIGHT=1.2 SIGMA=0.35,0.35
...
PRINT STRIDE=10 ARG=phi,psi,metad.bias FILE=COLVAR
\end{verbatim}
The input here is for a classic test case for these methods: calculating the free-energy surface for alanine dipeptide as a function of the $\phi$ and $\psi$ backbone dihedral angles.

What you have hopefully noticed from the above input files is that  PLUMED separates the calculation of the CVs from the calculation of the bias potential.  Consequently, when using PLUMED we would normally express the bias as: 
\begin{equation}
V(\vec{q},t) = V( s_1(\vec{q}), s_2(\vec{q}), \dots , s_n(\vec{q}), t) 
\end{equation}
This introduces a set of biased collective variables $\vec{s}(\vec{q}) = \{ s_1(\vec{q}), s_2(\vec{q}), \dots , s_n(\vec{q}) \}$ that allow us to represent the dynamics that all the atoms undergo in some low-dimensional space.  If we can calculate the partial derivatives of these collective variables with respect to the atomic positions we can then use the chain rule to calculate the forces on the atoms that result from the bias potential using:
\begin{equation}
f_m = - \frac{ \partial V }{ \partial q_m } = - \sum_{i=1}^n \frac{ \partial V }{\partial s_i } \frac{\partial s_i }{\partial q_m } 
\end{equation}
Here the sum runs over the $n$ collective variables that $V(\vec{s},t)$ is a function of.  $f_m$, meanwhile, is the force acting on one of the three components of the position of a particular atom,
and  $\frac{\partial s_i }{\partial x_m }$ is the derivative of the $i$th collective variable with respect to this particular component of the position of this same atom.

The key point to recognize is that the bias, which is some complicated function of the atomic positions, is expressed as a function of some set of simpler functions of the atomic positions.  Consequently, as we saw in the discussion of CVs, we can thus build very complicated bias potentials by combining simpler pieces together in the PLUMED input file. Notice also that, as discussed in Note \ref{note-stride}, we can make this business of computing bias potentials and CVs compatible with multiple-time-step algorithms for integrating the equations of motion.

As was the case for the CVs PLUMED contains implementations of other free-energy methods that we do not have the space to discuss here.  We will thus finish this section once more with a list of methods that are implemented in PLUMED together with relevant papers that describe them.  This list of methods includes:
 many variants on metadynamics
\cite{iannuzzi2003efficient,raiteri2006efficient,bussi2006free,piana2007bias,barducci2008well,branduardi2012metadynamics,dama2014well,dama2014transition,gil2015enhanced,pfaendtner2015efficient,hosek2016altruistic} including techniques that allow boundary conditions to be treated correctly in metadynamics simulations
\cite{baftizadeh2012protein} and
methods for extracting kinetic information from metadynamics simulations \cite{tiwary2013metadynamics}.  Then, in addition to metadynamics, we also have implementations of temperature-accelerated MD \cite{maragliano_ta,AbramsJ2008},
adaptive biasing force \cite{Lelievre2007,Zheng2012,Fu2016}, and
variationally enhanced sampling \cite{Valsson-PRL-2014,Valsson-JCTC-2015}.
Lastly, several bias potentials are included that allow one to force the ensemble averages that are extracted from the simulations to agree with data extracted from experiments.  The techniques for this that have been implemented include  experiment directed simulations \cite{white2014efficient,hocky2017cgds},
maximum entropy restraints \cite{cesari2016maxent},
target metadynamics \cite{white2015designing,marinelli2015ensemble,gil2016empirical},
and metainference \cite{Bonomi:2016ip}.

\subsection{Free Energies}
\label{sec:reweight-method}

To calculate the free energy as a function of the distance between two atoms using PLUMED and the technique described in section \ref{sec:histograms} one would use an input like the one shown below:
\begin{verbatim}
d1: DISTANCE ATOMS=1,2 
h: HISTOGRAM ...
   ARG=d1 GRID_MIN=0 GRID_MAX=1.0 
   GRID_BIN=200 KERNEL=DISCRETE STRIDE=10
...
f: CONVERT_TO_FES GRID=h TEMP=300
DUMPGRID GRID=f FILE=fes.dat   
\end{verbatim}
The input above creates a histogram with 200 bins that cover the range of distance values between 0 and 1~nm.  This histogram is constructed using the distance calculated for every 10th trajectory frame.  The free energy is then calculated from the accumulated histogram and output once the calculation has completed.

We can also use PLUMED and the methods described in section \ref{sec:histograms} to calculate the free-energy landscape from a biased simulation.  Below is a sample input that performs this type of calculation:
\begin{verbatim}
phi: TORSION ATOMS=1,2,3,4
psi: TORSION ATOMS=5,6,7,8
EXTERNAL ARG=phi,psi FILE=bias_potential.dat

as: REWEIGHT_BIAS TEMP=300

hh: HISTOGRAM ...
  LOGWEIGHTS=as ARG=d STRIDE=10 
  GRID_MIN=-pi,pi GRID_MAX=pi GRID_BIN=400 
  KERNEL=DISCRETE
... HISTOGRAM

ff: CONVERT_TO_FES GRID=hh TEMP=300
DUMPGRID GRID=ff FILE=fes.dat
\end{verbatim}
This is an input file for an enhanced sampling calculation that uses a bias that has been read from a file called \texttt{bias\_potential.dat}.  The bias potential that is read from this file is designed to act on the two torsional angles and to thus force them to sample configuration space more exhaustively.  The histogram is constructed using data from every 10th trajectory frame.  The free energy is then calculated from the accumulated histogram and output once the calculation has completed.  This is all very similar to what was done in the unbiased case.  The major difference here, however, is that we have used the \texttt{REWEIGHT\_BIAS} command and the \texttt{LOGWEIGHTS} keyword to ensure that the effect of the bias is discounted when the histogram is constructed.  The free-energy surface output is thus the free-energy surface for the unbiased Hamiltonian.

It is important to notice that weighting each of the simulated snapshots with a factor of $e^{+\frac{V(\vec{q})}{k_B T} }$
is correct when $V$ only depends on the coordinates of the systems. If the bias potential, $V$, is time-dependent,
more advanced procedures, that will not be discussed at length in this chapter, should be employed. Two particularly important enhanced sampling techniques that employ time dependent bias potential, and for which we provided sample inputs above, are:
\begin{itemize}
\item {\bf Steered MD}: To extract free energies from such simulations one has to run multiple steered MD calculations and then analyze the ensemble of trajectories obtained using the Jarzynki equality \cite{jarzynski1997nonequilibrium}. 
\item {\bf Metadynamics}: Several procedures have been proposed that allow one to compute appropriate weighting factors for these types of simulations \cite{bonomi2009reconstructing,branduardi2012metadynamics,tiwary2014time}.
\end{itemize}

\section{Calculating error bars}
\label{sec:errors}

The methods that we have introduced thus far for calculating ensemble averages and for calculating histograms are all based on the fact that we can calculate an approximate value for the true ensemble average by taking a large number of samples from the underlying distribution.  In other words, in all these methods we are calculating a sample mean and assuming that it provides a reasonable estimate for the true population mean.  What has thus far been missing from this discussion, however, is how we get an estimate for the error bar on the sample mean.  Any sample mean, after all, is itself a random variable and thus has an underlying distribution, which we should attempt to characterize.  The difficulty in doing so when our data is extracted from an MD trajectory is, however, that there are strong and measurable (see Note \ref{note-autocorrelation}) correlations between the configurations sampled in the frames of the trajectory that are adjacent.  As we will see in the remainder of this section, however, we can take these correlations into account when we calculate the variance and can thus calculate appropriate error bars.

\begin{figure}
\centering
\includegraphics[width=0.7\textwidth]{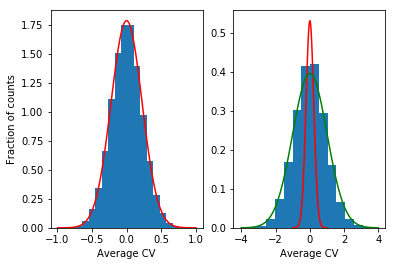}
\caption{Histograms that are obtained by sampling independent and identically distributed random variables (left) and by sampling CV values from a trajectory (right).  Each of the quantities that were used to calculate the histograms was constructed by taking an average of $n=20$ random variables.   The central limit theorem would thus predict the distribution of these averages to be a Gaussian with a variance of $\frac{\sigma^2}{n}$, where $\sigma^2$ is the sample variance.  The red lines thus show what this analytical function looks like for the two data sets.  For the independent random variables in the left panel it is clear that this is a good model for the sampled data.  For the correlated data in the right panel, however, this model substantially underestimates the variance in the true distribution of averages.  In fact, the green line shows a Gaussian that has the sample variance as its parameter and it is clear that this provides a much better model for the shape of this distribution.}
\label{fig:distribu}
\end{figure}

Figure~\ref{fig:distribu} shows why these correlations represent such a problem.  To construct the histogram shown in the panel on the left we generated 50000 sets of $n=20$ normal random variables.  We then calculated a sample average for each of these sets, $\mu_i$, and a sample mean, $\langle X\rangle$, and sample variance, $\sigma^2$, for all 1000000 points in the data set.  This sample variance was calculated using:
\begin{equation}
\sigma^2 = \frac{1}{N-1} \sum_{i=1}^N (X_i - \langle X \rangle)^2 
\end{equation}
The reason for doing this is that the central limit theorem tells us that the distribution of $\mu_i$ values should be a Gaussian centered on the sample mean with a variance of $\frac{\sigma^2}{n}$.  As you can see from the left panel this is a good model for the distribution of the sample means that was obtained by averaging the independent random variables.  The right panel shows, however, that this is not a good model when the data is taken from an MD or Monte Carlo trajectory.  As discussed in Note \ref{note-MC-model-data} to construct this figure sample means for 50000 sets of $n=20$ correlated random variables were calculated once more as well as the sample mean and variance for the full set of 1000000 variables.  The blue histogram shows the distribution of the sample means.  As you can see the correlation between the values of the CVs ensures that this distribution is substantially wider than the central limit theorem (red line) would predict it to be.  In other words, for correlated data $\frac{\sigma^2}{n}$ is not a sensible estimate for the variance of the sample mean.

This problem can be resolved using the so-called block averaging technique \cite{flyvbjerg1989error}.  In essence this technique involves splitting the trajectory into a set of $N$ blocks that are all of equal length.  $N$ separate sample means, $\mu_i$, are then calculated from each of these blocks of data.  Obviously, the mean for these $N$ quantities that is calculated using:
\begin{equation}
\mu = \frac{1}{N} \sum_{i=1}^N \mu_i
\end{equation}
is equal to the mean we would have obtained had we averaged every frame in our trajectory.  The advantage of block averaging in this way is, however, that, if each $\mu_i$ is calculated over a long enough block of trajectory, the values of the $\mu_i$s are no longer correlated.  We can thus calculate an error bar on our the estimate of $\mu$ that we would calculate using the formula above as:
\begin{equation}
\epsilon = \Phi^{-1}\left( \frac{p_c + 1}{2} \right) \sqrt{ \frac{1}{N(N-1)} \sum_{i=1}^N ( \mu_i - \mu )^2 }   
\end{equation}
where $\Phi^{-1}$ is the inverse of the cumulative probability distribution function for the standard normal random variable and where $p_c$ is the level of statistical confidence we want our error bar to represent. 

\begin{figure}
\centering
\includegraphics[width=0.8\textwidth]{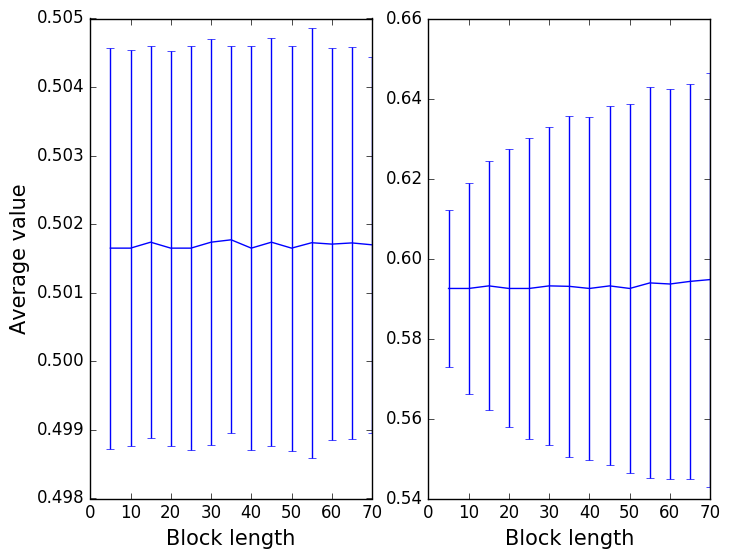}
\caption{Figure showing how the error bars calculated using the block averaging technique depend on the lengths of the blocks used for independent and identically distributed random variables (left) and for data from a typical MD/MC trajectory (right).  As you can see, the size of the error bar is largely independent of the block size when the data points being averaged are samples of a random variable.  The correlations between the CV values we obtain in a trajectory, however, ensure that the error bar is underestimated when the block size used is small.  Consequently, as the length of time over which the block averages are taken is increased the error bar increases in size until it eventually reaches a plateau.}
\label{fig:error-bars}
\end{figure}

Figure~\ref{fig:error-bars} shows the outcome of applying this technique on data obtained by sampling independent and identically distributed random variables and on data obtained by calculating the value of a CV over the course of a trajectory.  The mean and error bars are shown as a function of the sizes of the blocks over which the $\mu_i$ averages were calculated.  In both cases you can see that the final value of $\mu$ does not depend on the size of the blocks.  The small differences in this value are due to the finite precision algebra the computer uses.  For the correlated data from the MD trajectory, however, the size of the error bar does depend on the length of the blocks.  In particular, the error bar is smaller when the block sizes are small as the $\mu_i$ values are still correlated.  As the block sizes get larger, however, the $\mu_i$ values become decorrelated and the size of the error bar thus plateaus to a near constant value.  

A similar technique can be used to calculate error bars for histograms and free-energy surfaces.  For the weighted histograms discussed at the end of section \ref{sec:histograms} the probability, $g_j^{(i)}$, for the $j$th bin of the histogram is computed using equation \ref{eqn:weighted-histogram}, and the data from the $i$th block of trajectory data.  The following input shows how these histograms can be constructed in practise by using PLUMED.  
\begin{verbatim}
phi: TORSION ATOMS=1,2,3,4
psi: TORSION ATOMS=5,6,7,8
EXTERNAL ARG=phi,psi FILE=bias_potential.dat

as: REWEIGHT_BIAS TEMP=300

hh: HISTOGRAM ...
  LOGWEIGHTS=as ARG=d STRIDE=10 
  GRID_MIN=-pi,pi GRID_MAX=pi GRID_BIN=400 
  KERNEL=DISCRETE CLEAR=1000000
... HISTOGRAM

DUMPGRID GRID=hh FILE=histo.dat STRIDE=1000000
\end{verbatim}
The simulation that is being analyzed here is the same biased calculation that was described in  section \ref{sec:biasing}.  The input above, however, instructs PLUMED to output independent histograms from each block of 1,000,000 MD steps in the trajectory to a set of files called analysis.0.histo.dat, analysis.1.histo.dat, ... and histo.dat.  This procedure obviously gives multiple estimates for the probability of each bin, which can be combined to give a single set of sample means and confidence limits using the following or other similar expressions:
\begin{equation}
\langle g_j \rangle = \frac{1}{\sum_{i=1}^N w_i } \sum_{i=1}^N w_i g_j^{(i)}
\end{equation}
\begin{equation}
\label{eq:error-weights}
 \epsilon_j = \Phi^{-1}\left( \frac{p_c + 1}{2} \right) \sqrt{ \frac{ 1 }{N \left( W - \frac{S}{W} \right)} \sum_{i=1}^N w_i*( g_j^{(i)} - \langle g_j\rangle )^2 } 
\end{equation}
Here $w_i$ is used to indicate the sum of the weights of all the configurations in each of the blocks of data.  $W$ and $S$ are then the sum of these $w_i$ values and the sum of the squares of these values respectively.

\begin{figure}
\centering
\includegraphics[width=0.7\textwidth]{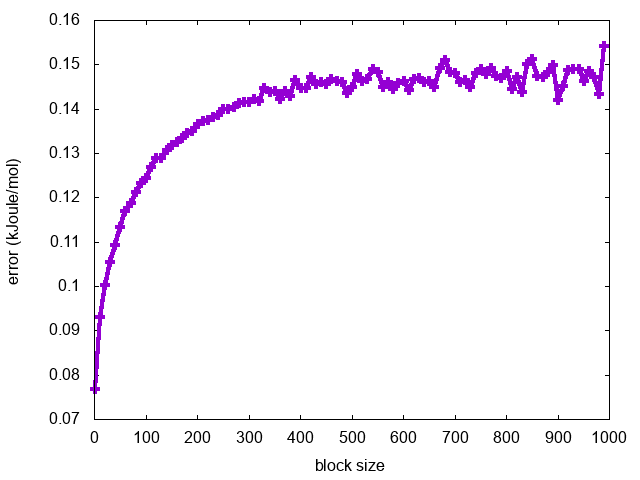}
\caption{Figure showing how the average error on the estimate of a free energy obtained for a metadynamics simulation of alanine dipeptide depends on the sizes of the blocks over which the individual histograms were calculated.  As with the example involving the ensemble average in figure \ref{fig:error-bars} the error is underestimated when small blocks are used.  When larger blocks of data are used, however, the value of the error plateaus at a constant value.}
\label{fig:fes-errors}
\end{figure}

Example python code for taking the set of histogram files output by PLUMED and for calculating the means and the error bars using the formulas above can be found on the PLUMED website.  Figure~\ref{fig:fes-errors} shows how the average size of the errors on the estimate of the free energy
obtained from a metadynamics simulation changes as the length of the blocks is increased
(see Note \ref{note-error-propagation}).
Similarly to what we observed in the right panel of figure \ref{fig:error-bars} when small blocks are used the error bar on the free energy is underestimated and the error bars are thus too small.  When sufficiently large blocks are used, however, the size of the error reaches a plateau value.

\section{Multiple replica techniques}
\label{sec:multiple-replicas}

Thus far we have considered cases where a single simulation is performed and then analyzed. There are, however, situations where one wants to simultaneously run multiple simulation replicas. Sometimes these replicas are run under equivalent conditions and the only differences between them are the initial configurations. If this is the case one can simply concatenate all the trajectories and analyze them using the techniques that have been discussed in the previous sections.  Alternatively, if the simulations are short and are thus not expected to reach equilibrium, more statistically reliable results may well be obtained if they are analyzed using the Markov state models that have been discussed in chapter IV. 
Oftentimes, however, the various replicas are run using different conditions (e.g., the replicas have different temperatures or different biasing potentials).
In this case the replicas will often also communicate with each other using a replica-exchange procedure.
In this section we will thus discuss how such multiple-replica simulations operate and how they should be analyzed. The analysis described in what follows can be applied when replicas are simulated independently of each other or when they are simulated in a manner that permits exchanges between replicas. For best results we would always suggest allowing exchanges between replicas as this increases the ergodicity of the simulation.

Replica-exchange molecular dynamics provides a rigorous theoretical framework that can be used to swap coordinates between trajectories that are calculated simultaneously but performed using different conditions (see Chapter II).
Perhaps the most commonly used of these techniques is parallel tempering \cite{sugita1999replica}, which uses a different temperature in each of the simulated replicas.  The replicas can also experience different simulation biases, however, so, because PLUMED is designed to bias MD simulations and because there are a number of techniques in the literature that are based on this principle (see, e.g., \cite{murata2004free,piana2007bias,curuksu2009enhanced,gil2015enhanced}), this will be the focus in this section.
Within many of these replica-exchange techniques the Metropolis Monte Carlo scheme is employed with proposed moves that involve swapping the coordinates of replica $i$ ($\vec{q}^{(i)}$) with the coordinates of replica $j$ ($\vec{q}^{(i)}$) and an acceptance probability that is given by:
\begin{equation}
\alpha = \min \left(
1,\frac{P^{(i)}(\vec{q}^{(j)})P^{(j)}(\vec{q}^{(i)})}{
P^{(i)}(\vec{q}^{(i)})P^{(j)}(\vec{q}^{(j)})}
\right)
\end{equation}
Here $P^{(i)}(\vec{q})$ is the probability that the set of coordinates $\vec{q}$ will be observed under the conditions experienced by the $i$th replica. 
Assuming that the only difference between the two replicas is the bias potential experienced and that the bias potentials for replica $i$ is $V^{(i)}(\vec{q})$ allows us to use equation \ref{eqn:biased-probability} to rewrite this probability using an expression that only depends on the bias potentials:
\begin{equation}
\alpha = \min \left(
1,\exp\left(\frac{
-V^{(i)}(\vec{q}^{(j)})
-V^{(j)}(\vec{q}^{(i)})
+V^{(i)}(\vec{q}^{(i)})
+V^{(j)}(\vec{q}^{(j)})
}{k_BT}\right)\right)
\end{equation}
When coordinate swaps are accepted or rejected using these criteria we ensure that the system remains at equilibrium. Consequently, any averages that we obtain are
in principle indistinguishable from those that would have been obtained if the simulations had been performed independently.  The advantage, however, is that the coordinate swaps usually systematically increase the number of slow transitions that are sampled. 
We can thus use these methods to obtain more statistically robust averages from shorter (parallel) simulations.

\subsection{The weighted histogram method}
\label{sec:wham}

The multiple replica simulations that were introduced above provide us with a powerful set of tools that allow us rapidly explore a wide region of configuration space and to take advantage of parallel computing facilities.
In this final section we will discuss how the trajectories we obtain from these simulations are analyzed.  Much like the analysis that was discussed in section \ref{sec:histograms} the aim here is to extract the free energy as a function of a CV or set of CVs.  Furthermore, when analyzing these multiple replica simulations we are again going to calculate a histogram and then exploit the maximum likelihood technique to convert this to a free energy.
 At odds with section \ref{sec:histograms}, however, we now have multiple trajectories, each of which was obtained by integrating a different biased Hamiltonian. We thus calculate the probability of observing this particular set of configurations during the $N$ trajectories that we ran using the product of multinomial distributions shown below:
\begin{equation}
P( \vec{T} ) \propto \prod_{j=1}^M \prod_{k=1}^N (c_k w_{kj} p_j)^{t_{kj}}
\label{eqn:wham1}
\end{equation}
In this expression the second product runs over the biases that were used when calculating the $N$ trajectories.  The first product then runs over the $M$ bins in our histogram.  The $p_j$ variable that is inside this product is the quantity we wish to extract; namely, the unbiased probability of having a set of CV values that lie within the range for the $j$th bin.
The quantity that we can easily extract from our simulations, $t_{kj}$, however, measures  the number of frames from trajectory $k$ that are inside the $j$th bin.  To interpret this quantity we must consider the bias that acts on each of the replicas so the $w_{kj}$ term is introduced.  This quantity is calculated using equation \ref{eqn:reweight} and is essentially the factor that we have to multiply the unbiased probability of being in the bin by in order to get the probability that we would be inside this same bin in the $k$th of our biased simulations.  Obviously, these $w_{kj}$ values depend on the value that the CVs take and also on the particular trajectory that we are investigating all of which, remember, have different simulation biases.  
Finally, $c_k$ is a free parameter that ensures that, for each $k$, the biased probability is normalized.

We can use equation \ref{eqn:wham1} to find a set of values for  $p_j$ that maximizes the likelihood of observing the trajectory. This constrained optimization must be performed using a set of Lagrange multipliers, $\lambda_k$, that ensure that each of the biased probability distributions are normalized, that is $\sum_j c_kw_{kj}p_j=1$.  Furthermore, much as in section \ref{sec:histograms}, the problem is made easier if equation \ref{eqn:wham1} is replaced by its logarithm.
\begin{equation}
\label{eq:lagrangian3}
\mathcal{L}= \sum_{j=1}^M \sum_{k=1}^N t_{kj} \ln c_k  w_{kj} p_j + \sum_k\lambda_k \left( \sum_{j=1}^N c_k w_{kj} p_j - 1 \right)
\end{equation}
After some manipulations (see Note \ref{note-lagrangian-wham} and the similar derivation that uses a slightly different notation that reported in Ref.\cite{bartels2000analyzing}), the following equations emerge:
\begin{equation}
\label{eqn:wham-f1}
p_j\propto \frac{\sum_{k=1}^N t_{kj}}{\sum_k c_k w_{kj}}
\end{equation}
\begin{equation}
c_k=\frac{1}{\sum_{j=1}^M w_{kj} p_j}
\label{eqn:wham-f2}
\end{equation}
Equations \ref{eqn:wham-f1} and \ref{eqn:wham-f2}
can be solved by computing the $p_j$ values using equation \ref{eqn:wham-f1} with an initial guess for the $c_k$ values and by then refining these $p_j$ values using the $c_k$ values that are obtained by inserting the $p_j$ values obtained into equation \ref{eqn:wham-f2}.  Usually the $c_k$ and $p_j$ values become self-consistent after a few rounds of this iterative algorithm.

When writing scripts to do this form of analysis it is worth noting that only $\sum_k t_{kj}$, which is the total number of configurations from \emph{all} the replicas that enter the $j$th bin, enters equations \ref{eqn:wham-f1} and \ref{eqn:wham-f2}.  There is thus no need to record which replica generated each of the frames and one can thus simply gather the trajectories from all the replicas together at the outset.

The fact that we can simply gather the trajectories from all our replicas before performing the weighted histogram analysis that has been described in the previous paragraph is also evident when we consider equation \ref{eqn:wham-f1}.  This expression tells us that we can calculate $p_j$ by constructing a weighted histogram from the concatenated trajectory and that the weight for the $n$th frame will be:
\begin{equation}
\tilde{w}_n = \frac{1}{\sum_k c_k w_{kn} }
\end{equation}
where the sum runs over the replicas and where $w_{kn}$ is the factor the $k$th replica has to be reweighted by in order to recover the unbiased probability for configuration $n$.  Notice, that we can also use the concatenated trajectory when extracting values for $c_k$ using equation \ref{eqn:wham-f2} as the sum over bins in the denominator can be replaced with a sum over the concatenated trajectory.  These observations are important as they are the basis of the binless formulation of the weighted histogram technique (WHAM) that is implemented within PLUMED and that has been variously proposed by a number of different authors
\cite{souaille2001extension,shirts2008statistically,tan2012theory}

\subsection{WHAM analysis with PLUMED}

\begin{figure}
\centering
\includegraphics[width=0.9\textwidth]{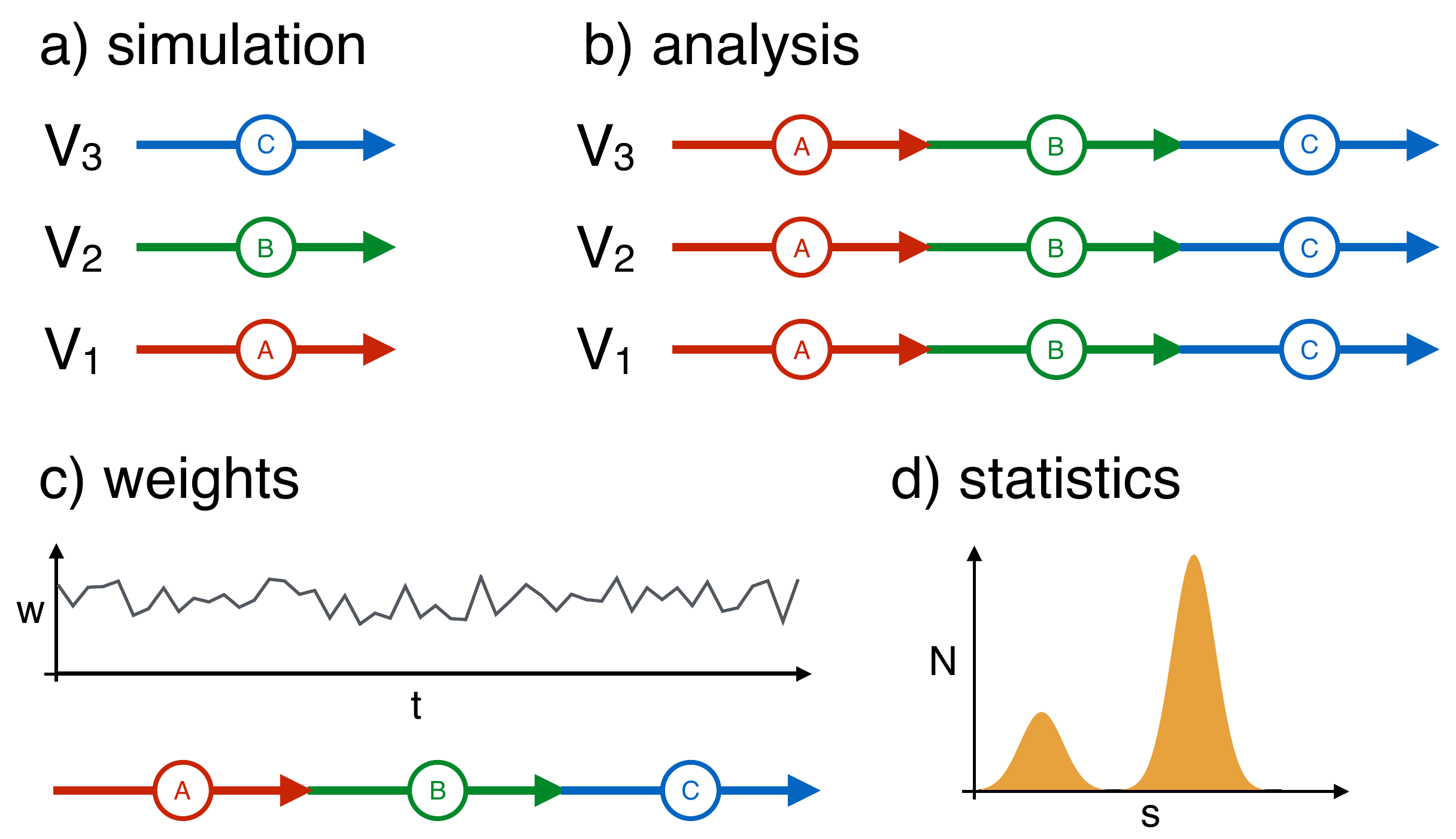}
\caption{Schematic representation of a WHAM analysis.
a) Multiple simulations are performed using different bias potentials. These simulations can be done separately, but we would always
recommend using a replica-exchange procedure. b) Trajectories are concatenated for subsequent analysis. The first step of this analysis is to compute the value of all the biasing potentials on all the snapshots in the concatenated trajectory.
Notice that in principle it is not necessary to concatenate the trajectory as the order of the accumulated snapshots is irrelevant.  It is critical, however, to compute all the bias potentials for all the frames in all the trajectories as only then can you solve equations \ref{eqn:wham-f1} and \ref{eqn:wham-f2}.
c) Once these equations are solved a weight for each snapshot in the concatenated trajectory is obtained. d) These weights are then used to, for example, reconstruct
the unbiased probability along some \emph{a posteriori} chosen CV.
}
\label{fig:wham}
\end{figure}

Section \ref{sec:multiple-replicas} discussed simulations in which multiple replicas are run in parallel and in which exchanges are attempted between replicas.  To run these types of calculations you need an MD code that can manage the communication between these replicas.  GROMACS is able to run simulations in this way using a command line syntax that is similar to the one shown below, which tells GROMACS to run 3 replicas in parallel and to attempt coordinate exchanges every 1000 MD steps:
\begin{verbatim}
> mpirun -np 3 gmx_mpi mdrun -multi 3 \
   -plumed plumed.dat -replex 1000
\end{verbatim}

For MD codes that can handle multiple replicas PLUMED provides a convenient syntax for having different biases on the various different replicas.  As an example the PLUMED input below assumes that three replicas are being run in parallel and that these three replicas differ only in the position of the center of the harmonic restraint:
\begin{verbatim}
d: DISTANCE ATOMS=10,20
RESTRAINT ARG=d AT=@replicas:1.0,1.1,1.2 KAPPA=1.0
\end{verbatim}
Obviously, the centers of the harmonic restraint in the three simulations are at 1.0, 1.1 and 1.2 respectively.  The CV on which this restraint acts and the strength of the restraint are, however, the same in all three replicas.

Once the multiple replica simulation has run, it must be analyzed.  As discussed in section \ref{sec:wham} the WHAM technique provides a good method for doing this analysis.  To do WHAM using PLUMED you must first \emph{concatenate} the trajectories from the various replicas.  The exact way this will be done will depend on the format of the trajectory file.
If the format is a plain text
\texttt{.gro} file, file concatenation may be sufficient.  For other file types, however, it may be necessary to use
the specific tools that are provided by the MD engine.  Regardless of these details, however, 
once a single concatenated trajectory is available it can be analyzed using PLUMED with an input file like the one shown below:
\begin{verbatim}
d: DISTANCE ATOMS=10,20
RESTRAINT ARG=d AT=@replicas:1.0,1.1,1.2 KAPPA=1.0

hh: WHAM_HISTOGRAM ...
   ARG=d BIAS=d.bias TEMP=300
   GRID_MIN=0 GRID_MAX=10 GRID_BIN=100 
   KERNEL=DISCRETE
...

ff: CONVERT_TO_FES GRID=hh TEMP=300
DUMPGRID GRID=ff FILE=fes.dat
\end{verbatim}
The first part of this input will be basically identical to the input used for the biased calculations. The rest, meanwhile, uses the functionality for reweighting that was discussed in section \ref{sec:reweight-method} through the shortcut command \texttt{WHAM\_HISTOGRAM} (see Note \ref{note-wham-histogram}). It is important to remember that there are multiple replicas when running the WHAM calculation using the input above as to deal with the replicas PLUMED \texttt{driver} has to be called using a command like that shown below:
\begin{verbatim}
> mpirun -np 3 plumed driver --multi 3 \
    --plumed plumed.dat
\end{verbatim}

We will now see how to use this syntax can be used to compute the free-energy landscape for an adenosine in water.
The simulation reported here was done using the setup and parameters described in Ref.~\cite{cesari2016maxent}.  Consequently, a simulation with 16 replicas was run using the command below:
\begin{verbatim}
> mpirun -np 16 gmx_mpi mdrun -multi 16 \
    -plumed plumed.dat -replex 1000
\end{verbatim}
with the following PLUMED input file
\begin{verbatim}
MOLINFO STRUCTURE=adenosine.pdb
chi: TORSION ATOMS=@chi-1
#
# Impose an umbrella potential on chi
# with a spring constant of 80 kjoule/mol
# and centered in chi=AT
#
r: RESTRAINT ...
  ARG=chi KAPPA=80.0
  AT=@replicas:{
  0*pi/8  1*pi/8  2*pi/8  3*pi/8
  4*pi/8  5*pi/8  6*pi/8  7*pi/8
  8*pi/8  9*pi/8  10*pi/8 11*pi/8
 12*pi/8 13*pi/8  14*pi/8 15*pi/8
 }
...
\end{verbatim}
The numbers listed after the \texttt{@replicas} instruction are basically 16 equally-spaced values between 0 and $2\pi$.  Furthermore, in setting up this grid of restraints we have exploited the fact
that PLUMED can perform simple algebraic calculations when interpreting its input.  

The restraints on the 16 replicas that will be simulated by executing the command above ensure that all the possible values for the $\chi$ glycosidic torsion of
this nucleoside, including the unfavorable ones that correspond to free-energy barriers, will be explored during these simulations.  Furthermore, once the simulation has completed we can concatenate all the trajectories produced by GROMACS
(called \texttt{traj0.xtc}, \texttt{traj1.xtc}, ... \texttt{traj15.xtc}) into a single long trajectory called \texttt{traj-all.xtc}.
This trajectory can then be analyzed using the following command
\begin{verbatim}
> mpirun -np 16 plumed driver --multi 16 \
   --plumed plumed.dat --ixtc traj-all.xtc
\end{verbatim}
and a PLUMED input file that is very similar to the one described above.
Instead of writing a new file from scratch, however, it is often more convenient to include the file that was used when the simulation was run into the analysis file by using the command \texttt{INCLUDE}. Doing so allows us to write an analysis file, called
\texttt{plumed\_wham.dat}, that reads as follows:
\begin{verbatim}
INCLUDE FILE=plumed.dat
# also compute the puckering of the sugar:
puck: PUCKERING ATOMS=@sugar-1

h1: WHAM_HISTOGRAM ...
   ARG=chi BIAS=r.bias TEMP=300
   GRID_MIN=-pi GRID_MAX=pi GRID_BIN=100 
   BANDWIDTH=0.1
...

fes1: CONVERT_TO_FES TEMP=300 GRID=h1
DUMPGRID GRID=fes1 FILE=fes1.dat

h2: WHAM_HISTOGRAM ...
   ARG=cc2.chi,cc2.puck.Zx BIAS=r.bias TEMP=300
   GRID_MIN=-pi,-pi GRID_MAX=pi,pi GRID_BIN=100,100 
   BANDWIDTH=0.1,0.1
...

fes2: CONVERT_TO_FES TEMP=300 GRID=h2
DUMPGRID GRID=fes2 FILE=fes2.dat
\end{verbatim}

\begin{figure}
\centering
\includegraphics[width=0.9\textwidth]{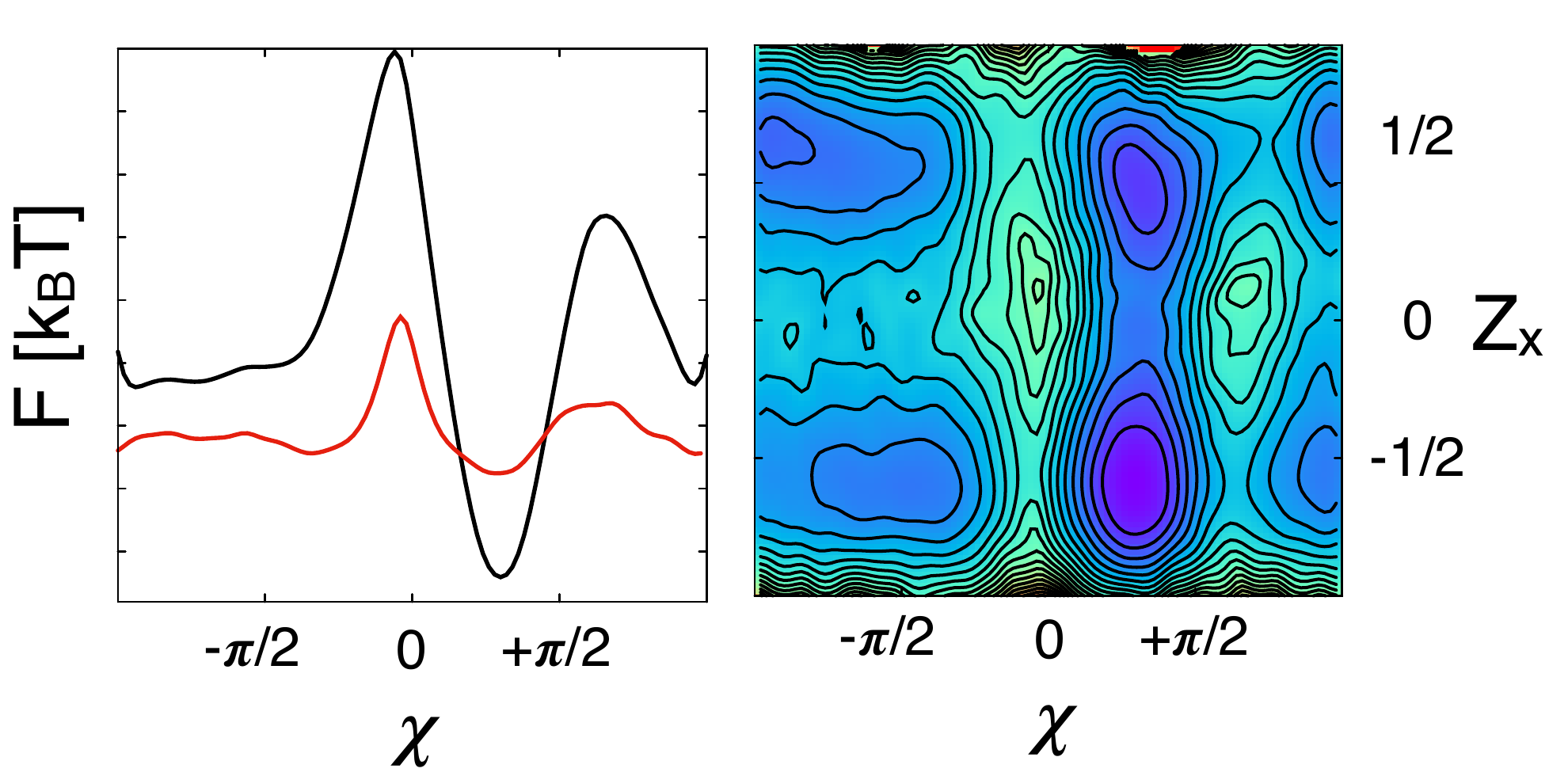}
\caption{
Results from a replica-exchange umbrella sampling simulation performed on adenosine in water.  As explained in the text restraints are applied on the glycosidic torsion $\chi$ in these simulations.  (a) The free energy as a function of $\chi$ computed using the WHAM equation is shown in black.  This profile is in agreement with that reported in reference \cite{cesari2016maxent}.
The free energy that one would have obtained if one had (erroneously) collected the histogram from the replica exchange simulation without performing any reweighting is also shown (red line). This profile is significantly flatter because the system is forced by the restraints to explore the entire CV span.
(b) Two-dimensional profile showing the free energy as a function of the biased $\chi$ torsion and the $Z_x$
puckering variable, that reports on the sugar conformation. Even though this latter variable was not biased, the
two-dimensional profile can be correctly reconstructed as transitions between the two typical conformations are accessible on the simulation timescale.
\label{fig:wham-A}
}
\end{figure}

For RNA, it is common to analyze the conformation of the sugar using the \texttt{Zx} component that is defined in \cite{huang2014improvement}.  To calculate this CV using PLUMED you use the \texttt{puck.Zx} component of the \texttt{PUCKERING} command.  
Notice that this CV is used in the above input because, in addition to computing the free energy as a function of the $\chi$ torsion, it also computes a second free-energy surface that depends on both $\chi$ and the
puckering conformation of the sugar.  This second variable was not biased, but it can also be analyzed at this latter stage.
The two free-energy surfaces that are extracted when the above analysis is performed on the trajectory are shown in figure \ref{fig:wham-A}.  In addition, this figure also shows the result of using a \texttt{HISTOGRAM} command \emph{without} taking the weighting factors into account.

Notice that the WHAM procedure discussed here can be used to remove any arbitrary bias that has been added to a simulation.
For instance, it can be used to analyze bias-exchange metadynamics simulations \cite{piana2007bias}, which is a method that involves using a replica-exchange scheme where each replica experiences a different metadynamics bias. In these types of calculations different replicas can use different PLUMED input files.  These replicas can even use a different number of biasing potentials.  For example, in references \cite{cunha2017unraveling,mlynsky2018molecular}, a similar procedure was used to reweight bias-exchange metadynamics simulations where, in addition to the metadynamics potential, each replica was subject to a different restraint.  In all these cases, however, the replicas always share the same simulation parameters.
In other words, the only differences are in the bias that is applied by PLUMED.

\section{Perspectives}

This chapter has introduced the PLUMED code, explained some of the theory behind the methods that are implemented in this code and given some practical examples that show how PLUMED can be used.  Space constraints mean that we cannot describe everything that PLUMED can do, so we have instead chosen to focus on some of the issues that come up most frequently on the code's mailing list.  The discussion in the previous sections has thus included information on the difficulties that can arise when treating periodic boundary conditions and a discussion on the proper treatment of statistical errors and the analysis of multiple replica simulations.  In what follows we will finish by giving a brief perspective on how we believe that PLUMED will evolve in the near future.

Our principal reason for developing PLUMED was to provide functionality for performing enhanced sampling calculations.  There were two interrelated reasons for writing this as a separate piece of software.  The first of these was that we wanted an inter-operable implementation that could work with multiple MD codes.  Different MD codes contain different functionalities and as such the particular engine that you will work with on any given project will depend on the particular system under study.  At the same time, however, the efficacy of any enhanced sampling method depends strongly on the degree to which the biased CVs separate the metastable basins and transition states in the energy landscape.  Consequently, PLUMED needed to be a large, stand-alone code not simply because some of the biasing methods that are implemented within are rather sophisticated but also because it contains implementations for many of the, in some cases very-complicated, CVs that have been used in these types of calculations.
In fact, to our knowledge there is no other code that contains implementations of as many different CVs.  Having said that, however, a number of recently developed CVs are not yet included and will thus be implemented in the near future.  We are particularly interested in some of the ideas from machine learning \cite{pamm,fieldcvs,tica-metad,ferg-auto-encoders,pande-auto-encoders} that are entering this field and are actively investigating how such methods could be implemented within the PLUMED framework.  Furthermore, we have in the last couple of years been working with the developers of Open Path Sampling \cite{ops} in order to help them interface their code with PLUMED so that methods such as transition path sampling and forward flux can be performed on-the-fly using a broad range of CVs without any need to perform a posteriori analysis.

Another issue that we need to work on in the future is the performance of the code.  As PLUMED is designed to compute and bias CVs its design is rather different from some of the other pieces of software that are used to drive MD simulations.  One big potential performance bottleneck comes about because PLUMED needs access to a subset of the simulated atoms during \emph{every time step.}  This requirement can slow down the simulations particularly when the CVs are computed from the positions of a large number of atoms. In the future we will thus work in order to decrease the computational cost of the operation that transfers the atomic positions from the MD code to PLUMED.  One easy way to reduce this cost, which will be available in version 2.5, is to allow expensive CVs that are computed within the underlying MD code to be transferred to PLUMED.  Computing the CV within the MD code would allow you take advantage of the data structures in the MD code.  Furthermore, as an added benefit, one could also take advantage of this feature when using CVs that are based on features that are difficult to transfer to PLUMED such as partial charges or other functions of the electronic structure. It is still to be seen if any these features will find a practical application, however.

One thing that will not change a great deal in the future is PLUMED's API.  The simple and well-documented interface between PLUMED and the various MD codes that call it is one of the code's great strengths so any future change will be made in a way that ensures backward compatibility.  The reason this interface is so important is that this is what allows such
a large number of MD codes to call PLUMED.
In addition, some developers have incorporated the PLUMED API within their code base so that users can download their codes and immediately use PLUMED without performing any sort of patching procedure.
This model of having the calls to PLUMED within the MD codes is something that we would like to use more widely in future.  Furthermore, to facilitate this, and to keep pace with the growing popularity of python within the molecular simulation community, we have recently provided an interface to the PLUMED API that makes the PLUMED routines callable from python.

The most important changes to PLUMED that we foresee, however,  will be  the features contributed to the code by independent groups.  It is clear from
 the github pages of PLUMED that a number of forks of the code are now being actively developed.
Furthermore, some of the features that have been developed in these forks have been already contributed back into the main version of the code. We are convinced that transforming this project in a community effort
will be the best way to keep it lively and up to date.  In fact, inviting contributions from the whole simulation community is the only way to ensure that the code  contains the most exciting recent methodological developments from the field. 

\section{Acknowledgements}

Writing and maintaining PLUMED involves a considerable amount of effort and we thus would like to finish by acknowledging everyone who has contributed to PLUMED in some way over the years.  PLUMED 2 was developed by a team of five core developers that includes the authors, Massimilliano Bonomi, Davide Branduardi and Carlo Camilloni.  Furthermore, Haochuan Chen, Haohao Fu, Glen Hocky, Omar Valsson, and Andrew White have all contributed modules to the code, whereas several other users have contributed other minor functionalities or fixed bugs in the code.  Lastly, we would like to acknowledge the many users and developers who have emailed our user and developer lists or attended the various PLUMED tutorials and user meetings.  The contributions of these people have been invaluable in terms of alerting us to bugs in the code.

\section{Notes}
\begin{enumerate}
\item
\label{note-plumed-versions}
The first version of PLUMED was released in 2009 \cite{bonomi2009plumed}.
A complete rewrite, that made the code easier to maintain and extend, was published in 2014 \cite{tribello2014plumed}. This second version changed the inner structure of the code and also changed the syntax for the input file so it is this second version of the code that will be the focus of this chapter. 
\item
\label{note-molinfo}
When a reference pdb file is provided using the MOLINFO command numerous shortcuts can be employed when calculating backbone torsional angles in proteins and nucleic acids.  For example the input below instructs PLUMED to calculate and print the $\phi$ and $\psi$ angles in the 3rd and 9th residue of a protein.
\begin{verbatim}
MOLINFO STRUCTURE=helix.pdb
phi3: TORSION ATOMS=@phi-3
psi3: TORSION ATOMS=@psi-3
phi9: TORSION ATOMS=@phi-9
psi9: TORSION ATOMS=@psi-9
PRINT ARG=phi3,psi3,phi9,psi9 FILE=colvar STRIDE=10
\end{verbatim}
A number of other convenient shortcuts are explained in the PLUMED manual.

\item
\label{note-rmsd}
If for some reason one only wishes to only disregard translation of the center of mass, that is to say if one wishes to include any displacements that come about because of rotation of the reference frame, when computing the RMSD one can replace \texttt{TYPE=OPTIMAL} with \texttt{TYPE=SIMPLE}. 
In addition, one can use one set of atoms to calculate the rototranslation operation that minimizes the RMSD and then use a different set of atoms to compute the final RMSD by adjusting the numbers that appear in the occupancy and beta columns of the input pdb file.  Using different sets of atoms to align the molecule and compute the RMSD displacement is commonly used when tracking the position of a ligand in the reference frame of a protein.
\item
\label{note-histogram}
The amount of time that the system spent in bin $j$ can be computed as follows:
\begin{equation}
t_j = \sum_{i=1}^T H\left( \frac{| s(\vec{q}_i) - s_j |}{ w } \right) \qquad \textrm{where} \qquad H(x) = 
\begin{cases} 
1 & \textrm{if} \quad  x< 1/2 \\
0 & \textrm{otherwise}
\end{cases}
\end{equation}
where $w$ is the width of each bin.
\item \label{note-lagrangian}
When we differentiate $\mathcal{L}$ in Eq.~\ref{eq:lagrangian1} with respect to $p_k$ we find that at the constrained optimum:
\begin{equation}
\frac{\partial \mathcal{L}}{\partial p_k} = \frac{t_k}{p_k} + \lambda = 0 \qquad \rightarrow \qquad p_k = - \frac{t_k}{\lambda} 
\end{equation}
We know, however, that:
\begin{equation}
\sum_{j=1}^M p_j = 1 \qquad \rightarrow \qquad \lambda = -\sum_{j=1}^M t_j 
\end{equation}
If we add together the number of trajectory frames in each of bins, however, we get the total number of trajectory frames, $T$.  $\lambda$ is thus equal to $-T$ and thus the most likely value of $p_k$ is simply:
\begin{equation}
p_k = \frac{t_k}{T}
\end{equation}
\item \label{note-lagrangian-weight}
When we differentiate $\mathcal{L}$ in Eq.~\ref{eq:lagrangian2} with respect to $p_k$ and set its derivatives to zero we obtain:
\begin{equation}
\frac{\partial \mathcal{L}}{\partial p_k} = \frac{t_j}{p_k} + w_k \lambda = 0 \qquad \rightarrow \qquad p_k = - \frac{(w_k)^{-1} t_k}{\lambda }
\end{equation}
We can then recall our constraint; namely $\sum_{j=1}^M w_j p_j = 1$.
Notice that by enforcing this constraint we are not doing anything to ensure that the unbiased distribution, $p_j$, is normalized. The constraint instead ensures that the biased distribution
$w_jp_j$ is normalized. This is important as it is this probability distribution that is used when we compute the total probability 
of observing the trajectory.
It is, in fact, not at all necessary for the unbiased probability distribution, $p_j$, to be normalized.  In fact, and to be clear, the unbiased distribution that emerges when this form of analysis is performed will be unnormalized as we will
obtain $\lambda=-T$ and hence $p_k=\frac{(w_k)^{-1}t_k}{T}$. The final result is thus:
\begin{equation}
p_k \propto (w_k)^{-1}t_k
\end{equation}
\item
\label{note-stride}

The \texttt{STRIDE} keyword takes a default value of one and tells you how frequently a PLUMED command is executed.
When it is used in combination with the \texttt{PRINT} command, it thus controls the frequency with which the CVs are printed. In addition, PLUMED automatically knows that these CVs
should only ever be calculated when they are printed. The \texttt{STRIDE} keyword can also be used with commands that bias CVs such as \texttt{RESTRAINT} and \texttt{METAD}, however. In this context the command tells PLUMED that the bias, and the biased CVs, should be computed with a frequency as part of a
a multiple-time step scheme \cite{tuckerman1992reversible,ferrarotti2014accurate}. By using these schemes you can speed
up calculations especially when expensive CVs are used.  You should, however, only ever use moderate values for \texttt{STRIDE} in this case -- typically, something between 1 and 5.

\item 
\label{note-autocorrelation}
We can measure the degree of correlation within a time series, $X_t$, of random variables with expectation $\langle X \rangle$ and variance $\langle (\delta X)^2 \rangle$ by measuring the autocorrelation function:
\begin{equation}
R(\tau) = \frac{ \langle ( X_t - \langle X \rangle ) ( X_{t+\tau} - \langle X \rangle ) \rangle}{ \langle (\delta X)^2 \rangle }
\end{equation}
The value of this function at $\tau$ gives a measure of the average degree of correlation between each pair of random variables that were measured $\tau$ time units apart.  If the random variables are all independent and identically distributed this function will decay to zero for all $t>0$.  If the autocorrelation function is calculated for a time series of CV values taken from a trajectory the function will not decay immediately, however, as 
the system will most likely not diffuse from one edge of CV space to the other during a single timestep. The CV value that we calculate from the $(i+1)$th frame of the trajectory will thus be similar to the value obtained for the $i$th trajectory frame.

\item 
\label{note-MC-model-data}

The Monte Carlo data that was used to construct the right panel of figure \ref{fig:distribu} was generated by performing a metropolis Monte Carlo simulation that sampled points from a standard normal distribution.

\item
\label{note-error-propagation}

The error bars, $\epsilon_j$, obtained for the components of the histogram, $\langle g_j\rangle$, that are calculated using equation \ref{eq:error-weights} must be propagated in order to obtain the error on the free energy. As the free energy is proportional to the logarithm of the probability, the error on this quantity is
$k_BT$
$\frac{\epsilon_j}{\langle g_j \rangle}$. 

\item
\label{note-lagrangian-wham}
When we differentiate $\mathcal{L}$ in Eq.~\ref{eq:lagrangian3} with respect to $p_k$ we find that at the constrained optimum:
\begin{equation}
\frac{\partial\mathcal{L}}{\partial p_j}= \sum_{k=1}^N \frac{t_{kj}}{ p_j} + \sum_{k=1}^N\lambda_k c_k w_{kj} = 0
\end{equation}
which can be rearranged to give:
\begin{equation}
\label{eq:wham-p}
p_j=-\frac{\sum_{k=1}^N t_{kj}}{\sum_k\lambda_k c_k w_{kj}}
\end{equation}
Similarly, when we differentiate $\mathcal{L}$ in Eq.~\ref{eq:lagrangian3} with respect to $c_k$ we find that
\begin{equation}
\frac{\partial\mathcal{L}}{\partial c_k}= \sum_{j=1}^M \frac{t_{kj}}{ c_k} + \sum_{j=1}^M\lambda_k w_{kj} p_j = 0
\end{equation}
which rearranges to give:
\begin{equation}
\lambda_k=-\frac{\sum_{j=1}^Mc_kw_{kj} p_j}{\sum_{j=1}^M t_{kj}}
\end{equation}
Finally, when we differentiate $\mathcal{L}$ in Eq.~\ref{eq:lagrangian3} with respect to $\lambda_k$ we
obtain the constraint:
\begin{equation}
\label{eq:wham-constraint}
\sum_{j=1}^M c_k w_{kj} p_j = 1
\end{equation}
The last two of these equations can be combined to give:
\begin{equation}
\label{eq:wham-lambda}
\lambda_k=-\frac{1}{\sum_{j=1}^M t_{kj}}
\end{equation}
Notice that $\sum_{j=1}^M t_{kj}$ is simply the length of trajectory $k$.
By assuming that all the trajectories have the same length we can thus ensure that  $\lambda_k$ is independent of $k$.
Inserting this result into equation \ref{eq:wham-p} will thus give:
\begin{equation}
p_j\propto \frac{\sum_{k=1}^N t_{kj}}{\sum_k c_k w_{kj}}
\end{equation}
Furthermore, rearranging equation \ref{eq:wham-constraint} gives:
\begin{equation}
c_k=\frac{1}{\sum_{j=1}^M w_{kj} p_j}
\end{equation}

\item
\label{note-wham-histogram}
When PLUMED reads in the command \texttt{WHAM\_HISTOGRAM} it converts it into the input for three actions automatically.  The first of these is a \texttt{REWEIGHT\_WHAM} command that is similar to the \texttt{REWEIGHT\_BIAS} command.  This command calculates a set of weights for the input configurations that are used when constructing weighted histograms using a \texttt{HISTOGRAM} command.  When using WHAM there is an important difference in computing the histogram, however.  When WHAM is used the weights can only be computed once the whole trajectory has been processed.  A special syntax and a \texttt{COLLECT\_FRAMES} command is thus required between the \texttt{REWEIGHT\_WHAM} and \texttt{HISTOGRAM} commands in this case. This special syntax thus instructs the action \texttt{HISTOGRAM} to wait until the end of the trajectory and to only then retrieve the weights and construct the histogram.  Notice also that equations \ref{eqn:wham-f1} and \ref{eqn:wham-f2} are solved in the \texttt{REWEIGHT\_WHAM} command and that this command can thus accept additional arguments in order to fine tune the tolerance with which these equations are solved.

\end{enumerate}


\begin{thebibliography}{10}
\providecommand{\url}[1]{{#1}}
\providecommand{\urlprefix}{URL }
\expandafter\ifx\csname urlstyle\endcsname\relax
  \providecommand{\doi}[1]{DOI \discretionary{}{}{}#1}\else
  \providecommand{\doi}{DOI \discretionary{}{}{}\begingroup
  \urlstyle{rm}\Url}\fi

\bibitem{bonomi2009plumed}
Bonomi M, Branduardi D, Bussi G, Camilloni C, Provasi D, Raiteri P, Donadio D,
  Marinelli F, Pietrucci F, Broglia RA, et~al. (2009), {PLUMED}: A portable
  plugin for free-energy calculations with molecular dynamics.
\newblock Comput Phys Commun  \textbf{180}(10), 1961

\bibitem{tribello2014plumed}
Tribello GA, Bonomi M, Branduardi D, Camilloni C, Bussi G (2014), {PLUMED} 2:
  New feathers for an old bird.
\newblock Comput Phys Commun  \textbf{185}(2), 604

\bibitem{abraham2015gromacs}
Abraham MJ, Murtola T, Schulz R, P{\'a}ll S, Smith JC, Hess B, Lindahl E
  (2015), {GROMACS}: High performance molecular simulations through multi-level
  parallelism from laptops to supercomputers.
\newblock SoftwareX  \textbf{1}, 19

\bibitem{plimpton1995fast}
Plimpton S (1995), Fast parallel algorithms for short-range molecular dynamics.
\newblock J Comput Phys  \textbf{117}(1), 1

\bibitem{todorov2006dl_poly_3}
Todorov IT, Smith W, Trachenko K, Dove MT (2006), {DL\_POLY\_3}: new dimensions
  in molecular dynamics simulations via massive parallelism.
\newblock J Mater Chem  \textbf{16}(20), 1911

\bibitem{hutter2014cp2k}
Hutter J, Iannuzzi M, Schiffmann F, VandeVondele J (2014), {CP2K}: atomistic
  simulations of condensed matter systems.
\newblock Wiley Interdiscip Rev Comput Mol Sci  \textbf{4}(1), 15

\bibitem{amber}
Case D, Betz R, Cerutti D, Cheatham T, III, Darden T, Duke R, Giese T,
  H.~Gohlke AG, Homeyer N, Izadi S, Janowski P, Kaus J, an AK, Lee T, LeGrand
  S, Li P, Lin C, Luchko T, Luo R, Madej B, Mermelstein D, Merz K, Monard G,
  Nguyen H, Nguyen H, Omelyan I, Onufriev A, Roe D, Roitberg A, Sagui C,
  Simmerling C, Botello-Smith W, J.~Swails RW, Wang J, Wolf R, Wu X, Xiao L,
  Kollman P.
\newblock {{AMBER} 2016, University of California, San Francisco} (2016)

\bibitem{eastman2017openmm}
Eastman P, Swails J, Chodera JD, McGibbon RT, Zhao Y, Beauchamp KA, Wang LP,
  Simmonett AC, Harrigan MP, Stern CD, et~al. (2017), {OpenMM} 7: rapid
  development of high performance algorithms for molecular dynamics.
\newblock PLOS Comput Biol  \textbf{13}(7), e1005659

\bibitem{fiorin2013using}
Fiorin G, Klein ML, H{\'e}nin J (2013), Using collective variables to drive
  molecular dynamics simulations.
\newblock Mol Phys  \textbf{111}(22-23), 3345

\bibitem{sidky2018ssages}
Sidky H, Col{\'o}n YJ, Helfferich J, Sikora BJ, Bezik C, Chu W, Giberti F, Guo
  AZ, Jiang X, Lequieu J, et~al. (2018), {SSAGES}: Software suite for advanced
  general ensemble simulations.
\newblock J Chem Phys  \textbf{148}(4), 044104

\bibitem{gil2015enhanced}
Gil-Ley A, Bussi G (2015), Enhanced conformational sampling using replica
  exchange with collective-variable tempering.
\newblock J Chem Theory Comput  \textbf{11}(3), 1077

\bibitem{best2013native}
Best RB, Hummer G, Eaton WA (2013), Native contacts determine protein folding
  mechanisms in atomistic simulations.
\newblock Proc Natl Acad Sci USA  \textbf{110}(44), 17874

\bibitem{camilloni2014statistical}
Camilloni C, Vendruscolo M (2014), Statistical mechanics of the denatured state
  of a protein using replica-averaged metadynamics.
\newblock J Am Chem Soc  \textbf{136}(25), 8982

\bibitem{zhang2011combined}
Zhang Y, Voth GA (2011), Combined metadynamics and umbrella sampling method for
  the calculation of ion permeation free energy profiles.
\newblock J Chem Theory Comput  \textbf{7}(7), 2277

\bibitem{de2016acidity}
De~Meyer T, Ensing B, Rogge SM, De~Clerck K, Meijer EJ, Van~Speybroeck V
  (2016), Acidity constant (pka) calculation of large solvated dye molecules:
  Evaluation of two advanced molecular dynamics methods.
\newblock ChemPhysChem  \textbf{17}(21), 3447

\bibitem{cheng2015solid}
Cheng B, Tribello GA, Ceriotti M (2015), Solid-liquid interfacial free energy
  out of equilibrium.
\newblock Phys Rev B  \textbf{92}(18), 180102

\bibitem{tribello2017clustering}
Tribello GA, Giberti F, Sosso GC, Salvalaglio M, Parrinello M (2017), Analyzing
  and driving cluster formation in atomistic simulations.
\newblock J Chem Theory Comput  \textbf{13}(3), 1317

\bibitem{peters2016reaction}
Peters B (2016), Reaction coordinates and mechanistic hypothesis tests.
\newblock Annu Rev Phys Chem  \textbf{67}, 669

\bibitem{kabsch1976solution}
Kabsch W (1976), A solution for the best rotation to relate two sets of
  vectors.
\newblock Acta Crystallogr A  \textbf{32}(5), 922

\bibitem{vymetal2011gyration}
Vymetal J, Vondrasek J (2011), Gyration-and inertia-tensor-based collective
  coordinates for metadynamics. application on the conformational behavior of
  polyalanine peptides and trp-cage folding.
\newblock J Phys Chem A  \textbf{115}(41), 11455

\bibitem{cunha2017unraveling}
Cunha RA, Bussi G (2017), Unraveling {Mg2+--RNA} binding with atomistic
  molecular dynamics.
\newblock RNA  \textbf{23}(5), 628

\bibitem{pietrucci2009collective}
Pietrucci F, Laio A (2009), A collective variable for the efficient exploration
  of protein beta-sheet structures: Application to {SH3} and {GB1}.
\newblock J Chem Theory Comput  \textbf{5}(9), 2197

\bibitem{bartels1998probability}
Bartels C, Karplus M (1998), Probability distributions for complex systems:
  adaptive umbrella sampling of the potential energy.
\newblock J Phys Chem B  \textbf{102}(5), 865

\bibitem{bonomi2010enhanced}
Bonomi M, Parrinello M (2010), Enhanced sampling in the well-tempered ensemble.
\newblock Phys Rev Lett  \textbf{104}(19), 190601

\bibitem{lazaridis1999effective}
Lazaridis T, Karplus M (1999), Effective energy function for proteins in
  solution.
\newblock Proteins  \textbf{35}(2), 133

\bibitem{do2013rna}
Do TN, Carloni P, Varani G, Bussi G (2013), {RNA}/peptide binding driven by
  electrostatics -- insight from bidirectional pulling simulations.
\newblock J Chem Theory Comput  \textbf{9}(3), 1720

\bibitem{nava2017dimer}
Nava M, Palazzesi F, Perego C, Parrinello M (2017), Dimer metadynamics.
\newblock J Chem Theory Comput  \textbf{13}(2), 425

\bibitem{bottaro2016free}
Bottaro S, Banas P, Sponer J, Bussi G (2016), Free energy landscape of {GAGA}
  and {UUCG} {RNA} tetraloops.
\newblock J Phys Chem Lett  \textbf{7}(20), 4032

\bibitem{spiwok2007metadynamics}
Spiwok V, Lipovov{\'a} P, Kr{\'a}lov{\'a} B (2007), Metadynamics in essential
  coordinates: free energy simulation of conformational changes.
\newblock J Phys Chem B  \textbf{111}(12), 3073

\bibitem{sutto2010comparing}
Sutto L, D’Abramo M, Gervasio FL (2010), Comparing the efficiency of biased
  and unbiased molecular dynamics in reconstructing the free energy landscape
  of met-enkephalin.
\newblock J Chem Theory Comput  \textbf{6}(12), 3640

\bibitem{branduardi2007b}
Branduardi D, Gervasio FL, Parrinello M (2007), From {A} to {B} in free energy
  space.
\newblock J Chem Phys  \textbf{126}(5), 054103

\bibitem{leines2012path}
Leines GD, Ensing B (2012), Path finding on high-dimensional free energy
  landscapes.
\newblock Phys Rev Lett  \textbf{109}(2), 020601

\bibitem{spiwok2011metadynamics}
Spiwok V, Kr{\'a}lov{\'a} B (2011), Metadynamics in the conformational space
  nonlinearly dimensionally reduced by isomap.
\newblock J Chem Phys  \textbf{135}(22), 224504

\bibitem{cremer1975general}
Cremer Dt, Pople J (1975), General definition of ring puckering coordinates.
\newblock J Am Chem Soc  \textbf{97}(6), 1354

\bibitem{huang2014improvement}
Huang M, Giese TJ, Lee TS, York DM (2014), Improvement of {DNA} and {RNA} sugar
  pucker profiles from semiempirical quantum methods.
\newblock J Chem Theory Comput  \textbf{10}(4), 1538

\bibitem{bonomi2017integrative}
Bonomi M, Camilloni C (2017), Integrative structural and dynamical biology with
  {PLUMED-ISDB}.
\newblock Bioinformatics  \textbf{33}(24), 3999

\bibitem{PCA_book}
Jolliffe I, \emph{Principal Component Analysis} (Springer, 2002)

\bibitem{MDS_book}
Borg I, Groenen PJF, \emph{Modern Multidimensional Scaling: theory and
  applications} (Springer, 2005)

\bibitem{sketchmap}
Ceriotti M, Tribello GA, Parrinello M (2011), Simplifying the representation of
  complex free-energy landscapes using sketch-map.
\newblock Proc Natl Acad Sci USA  \textbf{108}(32), 13023

\bibitem{giorgino2014plumed}
Giorgino T (2014), {PLUMED-GUI}: An environment for the interactive development
  of molecular dynamics analysis and biasing scripts.
\newblock Comput Phys Commun  \textbf{185}(3), 1109

\bibitem{humphrey1996vmd}
Humphrey W, Dalke A, Schulten K (1996), {VMD}: visual molecular dynamics.
\newblock J Mol Graph  \textbf{14}(1), 33

\bibitem{torrie1977nonphysical}
Torrie GM, Valleau JP (1977), Nonphysical sampling distributions in monte carlo
  free-energy estimation: Umbrella sampling.
\newblock J Comput Phys  \textbf{23}(2), 187

\bibitem{kumar1992weighted}
Kumar S, Rosenberg JM, Bouzida D, Swendsen RH, Kollman PA (1992), The weighted
  histogram analysis method for free-energy calculations on biomolecules. {I}.
  {T}he method.
\newblock J Comput Chem  \textbf{13}(8), 1011

\bibitem{isralewitz1997binding}
Isralewitz B, Izrailev S, Schulten K (1997), Binding pathway of retinal to
  bacterio-opsin: a prediction by molecular dynamics simulations.
\newblock Biophys J  \textbf{73}(6), 2972

\bibitem{laio2002escaping}
Laio A, Parrinello M (2002), Escaping free-energy minima.
\newblock Proc Natl Acad Sci USA  \textbf{99}(20), 12562

\bibitem{iannuzzi2003efficient}
Iannuzzi M, Laio A, Parrinello M (2003), Efficient exploration of reactive
  potential energy surfaces using {Car-Parrinello} molecular dynamics.
\newblock Phys Rev Lett  \textbf{90}(23), 238302

\bibitem{raiteri2006efficient}
Raiteri P, Laio A, Gervasio FL, Micheletti C, Parrinello M (2006), Efficient
  reconstruction of complex free energy landscapes by multiple walkers
  metadynamics.
\newblock J Phys Chem B  \textbf{110}(8), 3533

\bibitem{bussi2006free}
Bussi G, Gervasio FL, Laio A, Parrinello M (2006), Free-energy landscape for
  $\beta$ hairpin folding from combined parallel tempering and metadynamics.
\newblock J Am Chem Soc  \textbf{128}(41), 13435

\bibitem{piana2007bias}
Piana S, Laio A (2007), A bias-exchange approach to protein folding.
\newblock J Phys Chem B  \textbf{111}(17), 4553

\bibitem{barducci2008well}
Barducci A, Bussi G, Parrinello M (2008), Well-tempered metadynamics: a
  smoothly converging and tunable free-energy method.
\newblock Phys Rev Lett  \textbf{100}(2), 020603

\bibitem{branduardi2012metadynamics}
Branduardi D, Bussi G, Parrinello M (2012), Metadynamics with adaptive
  {Gaussians}.
\newblock J Chem Theory Comput  \textbf{8}(7), 2247

\bibitem{dama2014well}
Dama JF, Parrinello M, Voth GA (2014), Well-tempered metadynamics converges
  asymptotically.
\newblock Phys Rev Lett  \textbf{112}(24), 240602

\bibitem{dama2014transition}
Dama JF, Rotskoff G, Parrinello M, Voth GA (2014), Transition-tempered
  metadynamics: robust, convergent metadynamics via on-the-fly transition
  barrier estimation.
\newblock J Chem Theory Comput  \textbf{10}(9), 3626

\bibitem{pfaendtner2015efficient}
Pfaendtner J, Bonomi M (2015), Efficient sampling of high-dimensional
  free-energy landscapes with parallel bias metadynamics.
\newblock J Chem Theory Comput  \textbf{11}(11), 5062

\bibitem{hosek2016altruistic}
Hosek P, Toulcova D, Bortolato A, Spiwok V (2016), Altruistic metadynamics:
  Multisystem biased simulation.
\newblock J Phys Chem B  \textbf{120}(9), 2209

\bibitem{baftizadeh2012protein}
Baftizadeh F, Cossio P, Pietrucci F, Laio A (2012), Protein folding and
  ligand-enzyme binding from bias-exchange metadynamics simulations.
\newblock Curr Phys Chem  \textbf{2}(1), 79

\bibitem{tiwary2013metadynamics}
Tiwary P, Parrinello M (2013), From metadynamics to dynamics.
\newblock Phys Rev Lett  \textbf{111}(23), 230602

\bibitem{maragliano_ta}
Maragliano L, Vanden-Eijnden E (2006), A temperature accelerated method for
  sampling free energy and determining reaction pathways in rare events
  simulations.
\newblock Chem Phys Lett  \textbf{426}, 168

\bibitem{AbramsJ2008}
Abrams JB, Tuckerman ME ({2008}), {Efficient and Direct Generation of
  Multidimensional Free Energy Surfaces via Adiabatic Dynamics without
  Coordinate Transformations}.
\newblock J Phys Chem B  \textbf{{112}}({49}), 15742

\bibitem{Lelievre2007}
Leli{\`{e}}vre T, Rousset M, Stoltz G (2007), Computation of free energy
  profiles with parallel adaptive dynamics.
\newblock J Chem Phys  \textbf{126}(13), 134111

\bibitem{Zheng2012}
Zheng L, Yang W (2012), Practically efficient and robust free energy
  calculations: Double-integration orthogonal space tempering.
\newblock J Chem Theory Comput  \textbf{8}(3), 810

\bibitem{Fu2016}
Fu H, Shao X, Chipot C, Cai W (2016), Extended adaptive biasing force
  algorithm. {A}n on-the-fly implementation for accurate free-energy
  calculations.
\newblock J Chem Theory Comput  \textbf{12}(8), 3506

\bibitem{Valsson-PRL-2014}
Valsson O, Parrinello M (2014), Variational approach to enhanced sampling and
  free energy calculations.
\newblock Phys Rev Lett  \textbf{113}(9), 090601

\bibitem{Valsson-JCTC-2015}
Valsson O, Parrinello M (2015), {Well-Tempered Variational Approach to Enhanced
  Sampling}.
\newblock J Chem Theory Comput  \textbf{11}(5), 1996

\bibitem{white2014efficient}
White AD, Voth GA (2014), {An Efficient and Minimal Method to Bias Molecular
  Simulations with Experimental Data}.
\newblock J Chem Theory Comput  \textbf{10}, 3023

\bibitem{hocky2017cgds}
Hocky GM, Dannenhoffer-Lafage T, Voth GA (2017), Coarse-grained directed
  simulation.
\newblock J Chem Theory Comput  \textbf{13}(9), 4593

\bibitem{cesari2016maxent}
Cesari A, Gil-Ley A, Bussi G (2016), Combining simulations and solution
  experiments as a paradigm for {RNA} force field refinement.
\newblock J Chem Theory Comput  \textbf{12}(12), 6192

\bibitem{white2015designing}
White AD, Dama JF, Voth GA (2015), Designing free energy surfaces that match
  experimental data with metadynamics.
\newblock J Chem Theory Comput  \textbf{11}(6), 2451

\bibitem{marinelli2015ensemble}
Marinelli F, Faraldo-G{\'o}mez JD (2015), Ensemble-biased metadynamics: A
  molecular simulation method to sample experimental distributions.
\newblock Biophys J  \textbf{108}(12), 2779

\bibitem{gil2016empirical}
Gil-Ley A, Bottaro S, Bussi G (2016), Empirical corrections to the amber {RNA}
  force field with target metadynamics.
\newblock J Chem Theory Comput  \textbf{12}(6), 2790

\bibitem{Bonomi:2016ip}
Bonomi M, Camilloni C, Cavalli A, Vendruscolo M (2016), {Metainference: A
  Bayesian inference method for heterogeneous systems.}
\newblock Sci Adv  \textbf{2}(1), e1501177

\bibitem{jarzynski1997nonequilibrium}
Jarzynski C (1997), Nonequilibrium equality for free energy differences.
\newblock Phys Rev Lett  \textbf{78}(14), 2690

\bibitem{bonomi2009reconstructing}
Bonomi M, Barducci A, Parrinello M (2009), Reconstructing the equilibrium
  {Boltzmann} distribution from well-tempered metadynamics.
\newblock J Comput Chem  \textbf{30}(11), 1615

\bibitem{tiwary2014time}
Tiwary P, Parrinello M (2014), A time-independent free energy estimator for
  metadynamics.
\newblock J Phys Chem B  \textbf{119}(3), 736

\bibitem{flyvbjerg1989error}
Flyvbjerg H, Petersen H (1989), Error estimates on averages of correlated data.
\newblock J Chem Phys  \textbf{91}(1), 461

\bibitem{sugita1999replica}
Sugita Y, Okamoto Y (1999), Replica-exchange molecular dynamics method for
  protein folding.
\newblock Chem Phys Lett  \textbf{314}(1), 141

\bibitem{murata2004free}
Murata K, Sugita Y, Okamoto Y (2004), Free energy calculations for {DNA} base
  stacking by replica-exchange umbrella sampling.
\newblock Chem Phys Lett  \textbf{385}(1), 1

\bibitem{curuksu2009enhanced}
Curuksu J, Zacharias M (2009), Enhanced conformational sampling of nucleic
  acids by a new hamiltonian replica exchange molecular dynamics approach.
\newblock J Chem Phys  \textbf{130}(10), 03B610

\bibitem{bartels2000analyzing}
Bartels C (2000), Analyzing biased {Monte} {Carlo} and molecular dynamics
  simulations.
\newblock Chem Phys Lett  \textbf{331}(5-6), 446

\bibitem{souaille2001extension}
Souaille M, Roux B (2001), Extension to the weighted histogram analysis method:
  combining umbrella sampling with free energy calculations.
\newblock Comput Phys Commun  \textbf{135}(1), 40

\bibitem{shirts2008statistically}
Shirts MR, Chodera JD (2008), Statistically optimal analysis of samples from
  multiple equilibrium states.
\newblock J Chem Phys  \textbf{129}(12), 124105

\bibitem{tan2012theory}
Tan Z, Gallicchio E, Lapelosa M, Levy RM (2012), Theory of binless multi-state
  free energy estimation with applications to protein-ligand binding.
\newblock J Chem Phys  \textbf{136}(14), 04B608

\bibitem{mlynsky2018molecular}
Ml{\`y}nsk{\`y} V, Bussi G, et~al. (2018), Molecular dynamics simulations
  reveal an interplay between {SHAPE} reagent binding and {RNA} flexibility.
\newblock J Phys Chem Lett  \textbf{9}, 313

\bibitem{pamm}
Gasparotto P, Ceriotti M (2014), Recognizing molecular patterns by machine
  learning: An agnostic structural definition of the hydrogen bond.
\newblock J Chem Phys  \textbf{141}(17), 174110

\bibitem{fieldcvs}
Tribello GA, Ceriotti M, Parrinello M (2012), Using sketch-map coordinates to
  analyze and bias molecular dynamics simulations.
\newblock Proc Natl Acad Sci USA  \textbf{109}(14), 5196

\bibitem{tica-metad}
M.~Sultan M, Pande VS (2017), {TICA}-metadynamics: Accelerating metadynamics by
  using kinetically selected collective variables.
\newblock J Chem Theory Comput  \textbf{13}(6), 2440

\bibitem{ferg-auto-encoders}
{Chen} W, {Ferguson} AL (2018), {Molecular enhanced sampling with autoencoders:
  On-the-fly collective variable discovery and accelerated free energy
  landscape exploration}  ArXiv:1801.00203

\bibitem{pande-auto-encoders}
{Sultan} MM, {Wayment-Steele} HK, {Pande} VS (2018), {Transferable neural
  networks for enhanced sampling of protein dynamics}.
\newblock J Chem Theory Comput  \textbf{14}(4), 1887

\bibitem{ops}
Open path sampling.
\newblock http://openpathsampling.org/latest

\bibitem{tuckerman1992reversible}
Tuckerman M, Berne BJ, Martyna GJ (1992), Reversible multiple time scale
  molecular dynamics.
\newblock J Chem Phys  \textbf{97}(3), 1990

\bibitem{ferrarotti2014accurate}
Ferrarotti MJ, Bottaro S, P{\'e}rez-Villa A, Bussi G (2014), Accurate multiple
  time step in biased molecular simulations.
\newblock J Chem Theory Comput  \textbf{11}(1), 139

\end{thebibliography}
\end{document}